# Determination of the Zak phase of one-dimensional photonic systems via far-field radiations in Fourier space

C. Liu[1], H.R. Wang[1], and H.C. Ong[2]

Department of Physics, The Chinese University of Hong Kong, Shatin, Hong Kong, People's Republic of China

Bloch waves in 1D periodic systems carry Zak phase, which plays a key role in determining the band topology. In general, for a system that possesses inversion symmetry, the Zak phase of an isolated band is quantized as 0 or $\pi$ and is associated with the spatial field symmetries of the Bloch waves at the Brillouin zone center and boundary. Since the radiation losses from leaky systems are strongly associated with the Bloch waves, one may probe the far-field continuum to determine the Zak phases. Here, we formulate the radiations from photonic systems in Fourier space at the zone center and boundary and find they reveal the field symmetries and thereby the corresponding Zak phase. For verification, we have studied the Zak phases of TM plasmonic and TE photonic crystals by electrodynamic simulations and measuring the topological properties of plasmonic crystals using Fourier space optical spectroscopy and common path interferometry. In addition, a topological protected interface state is demonstrated when two topological trivial and nontrivial systems are joined together. Our results provide a general way for characterizing the band topology of photonic systems via far-fields.

1) These authors equally contributed to this work
2) hcong@phy.cuhk.edu.hk

## I. INTRODUCTION

Topological physics has attracted a widespread of interest not only in condensed matter physics [1-3] but also in other branches such as ultracold atom [4,5], electromagnetism [6-8], mechanics [9], acoustics [10,11], and oceanography [12]. Much attention in this field focuses on realizing the so-called topologically protected states, which support robust wave propagation against perturbation and disorder [1-12]. To produce such states, two systems that are topologically trivial and nontrivial are brought together to facilitate the occurrence of topological phase transition at the interface. As most of the matters are topologically trivial, the identification and the making of different classes of topological systems are currently under intensive investigation [13,14]. Likewise, developing methods to characterize the topological properties of the systems is also of great importance.

In analogy to the Su-Schrieffer-Heeger (SSH) model, the band topology of a 1D periodic systems is determined by Zak phase, $\gamma$, which is a geometric phase [15,16]. For the $\ell^{th}$ isolated energy band, the $\gamma_\ell$ emerges when the Bloch wave travels adiabatically along the band across the first Brillouin zone from $k = -\pi/P$ to $\pi/P$, where P is the period of the system. [16]. If the system possesses inversion symmetry, the $\gamma_\ell$ is quantized as either 0 or $\pi$ [16]. Zero or $\pi$ $\gamma$ defines the topological invariant of a two-band system. For a system that supports multiple bands, the topology of the band gap of interest is the summation of all $\gamma$ of the bands below that gap, giving rise to the total $\gamma$ that is either even or odd multiple of $\pi$ for indicating whether the system is topologically trivial or nontrivial [17,18]. A zero-dimensional interface state is then formed between two odd and even $\pi$ systems.

One notable feature that comes with $\gamma$ is the distinctive spatial wave symmetries at the zone center and boundary of the band [16-18]. The field symmetries are the same for $\gamma = 0$ but different when γ = π [18]. The association between γ and the field symmetry can be understood from the standpoint of Wannier function, which sums the Bloch waves carrying all k along a band [19]. Consider the Bloch waves at two high symmetry points that have the same field symmetry, the Wannier function has $W(-x) = \pm W(x)$ spatial dependence, leading to γ = $\frac{2\pi}{P}\int_{-\infty}^{\infty} x|W(x)|^2 dx$ = 0 [16]. On the other hand, for the waves that exhibit different spatial symmetries, the Wannier function now shows $W(-x+P) = \pm W(x)$ dependence, which gives γ = π [16]. Remarkably, the quantization of γ and the associated field symmetries are carried over to non-Hermitian systems provided they still possess inversion symmetry [20]. Therefore, instead of tracing the Bloch waves one by one along the band to determine γ, one can simply examine the field symmetries. However, how to measure the spatial wave symmetry remains challenging.

To date, there have been a few studies on measuring the geometric phase, either Zak or Berry phase [21-24]. While most of them focus on tracing the Bloch waves in momentum space, Gorlach et al adopt an alternative approach by probing the spectral positions of the dipolar (bright) and quadrupolar (dark) supermodes at the zone center, which reveal the topological invariant of the system [25]. However, their method is limited to 2D honeycomb systems that possess $C_6$ symmetry in which up and down pseudo-spins are found to define the nonzero spin Chern number [26]. Therefore, measuring the supermodes only at the zone center is sufficient to determine the band topology. In addition, it is applicable only to the lowest band gap above the light line but not others at higher energy. Recently, Chan and his coworkers have formulated that the sign of the reflection phase for wavelengths within the $\ell^{th}$ band gap can resolves the $\gamma_\ell$ of 1D systems [17,18]. Although the approach has been applied to several photonic and acoustic systems, it is demonstrated specifically to stacked multilayer structures which support one single normal reflection channel for phase measurement [27-30]. Particularly, for photonic systems, a stringent interferometric configuration is required that may be difficult to be implemented to other more general systems such as corrugated surfaces [27].

Here, we have developed a simple approach to determine γ by measuring the radiations from photonic systems in Fourier space at the Brillouin zone center and boundary. It is shown the radiations emitted by the supermodes exhibit a rich variety of characteristics such as quasi bound state in the continuum (BIC) [31,32] and even and odd mirror symmetric diffraction pairs that reveal the near-field symmetry. To verify our formulation, we perform finite-difference time-domain (FDTD) simulations on 1D Au plasmonic and $SiO_2$/Au photonic crystals that respectively support TM- and TE-polarized guided waves. We then fabricate plasmonic crystals (PmCs) with different geometries and conduct polarization- and angle-resolved diffraction and phase spectroscopy by Fourier space optical microscopy and common path interferometry to study γ. Both the simulation and experimental results agree very well with the theory. Finally, a topological protected interface state is demonstrated by joining two topological trivial and nontrivial PmCs together.

## II.  TEMPORAL COUPLED MODE THEORY

Assume a 1D optically thick periodic system that supports guided Bloch modes in the Γ-X direction is excited by light illuminated from the reflection side, Fig. 1(a) illustrates the TM- or TE-polarized ω-k plot with the in-plane wavevector k = $2\pi \sin\theta_i/\lambda$, where $\theta_i$ is the incident angle [33-36]. The plot is also known as dispersion relation, and it provides the necessary conditions for exciting the resonant modes (solid lines) in the system. In analogous to the band folding scheme in electronic crystals, the excitation requires the incident light to fulfill the phase-matching equation given as $\left(k + n_B \frac{2\pi}{P}\right)^2 = \left(n_{eff}\frac{\omega}{c}\right)^2$, where $n_B$ is the Bragg scattering order, c is the light speed, and P and $n_{eff}$ are the period and the wavelength-dependent effective refractive index of the system, in analogy to the band folding scheme in electronic crystals [33,35,36]. Depending on $n_B$, the modes form several continuous energy bands (solid lines) spanning across the first Brillouin zone and are divided by band gaps at the Γ and X points. The gap arises when two degenerate counter-propagating Bloch modes interact to yield two non-propagating supermodes locating at higher and lower energies [33,34]. Because the $n_{eff}$ of the system is larger than 1, the spectral positions of the supermodes are lower than the cross-points of the Wood's anomalies (dashed lines) where $n_{eff}$ is taken to be 1 [37]. The cross-points at the Γ and X points are positioned at $\lambda_o = \frac{P}{M}$ and $\frac{P}{M-1/2}$, respectively, with M is a nonzero positive integer that increases sequentially with the cross-point position as shown in Fig. 1(a) [38]. In this study, we will focus on the band sandwiched between the second and third gaps with the supermodes of interest marked as two blue circles. The generalization of other higher order bands will be provided in the Supplementary Information [38]. Given the system possesses inversion symmetry, the near-field symmetries of the supermodes sitting on the band thus define the γ to be 0 or π.

As the system is leaky, once the supermode is excited, it will then dissipate into discrete radiation channels via diffraction. In fact, the output channels in Fourier space follow the grating equation given as $m\lambda = P(sin\theta_i + sin\theta_m)$, where $\theta_m$ is the diffraction angle with m is the diffraction order [39]. For example, for the second band gap at the Γ point where M = 1 and $\lambda > \lambda_o = P$, the circled supermode supports only one single m = 0 specular diffraction order at $\theta_{m=0} = 0°$, as shown in Fig. 1(b). Other nonzero m integers do not yield propagating orders in the continuum. Likewise, for the third gap at k = π/P in Fig. 1(c) where M = 2 and $\lambda > \lambda_o = 2P/3$, we see $\frac{2m-1}{3}\frac{\lambda}{\lambda_o} = sin\theta_m$ results in m = 0 and 1 as the possible solutions, and $(2m-1)sin\theta_i = sin\theta_m$ shows two diffractions occur at $\theta_{m=0,1} = \pm\theta_i$. These two specular and back reflected orders are mirror symmetric with respect to the surface normal of the system. Their sum gives the total reflection. In the Supplementary Information, we will extend the diffraction equation to higher band gaps and show the number of channels from each supermode at the Γ and X points are 2(M-1)+1 and 2(M-1) [38]. In addition, the 2(M-1) channels at two points are always presented as mirror symmetric pairs. However, the normal diffraction order is present at the Γ point but is absent at the X point.

The interaction between two degenerate Bloch modes and the far-field channels can be formulated within the framework of CMT [40-42]. For a lossy and leaky system, at either the Γ or X point, the dynamics of two counter-propagating Bloch mode amplitudes, $a_1$ and $a_2$, taken under TM or TE polarization are written as:

$$\frac{d}{dt}\begin{bmatrix} a_1 \\ a_2 \end{bmatrix} = i\begin{bmatrix} \tilde{\omega}_o & \tilde{\omega}_c \\ \tilde{\omega}_c & \tilde{\omega}_o \end{bmatrix}\begin{bmatrix} a_1 \\ a_2 \end{bmatrix} + K^T\begin{bmatrix} s_{m,+} \end{bmatrix}, \tag{1}$$

where $\tilde{\omega}_o$ and $\tilde{\omega}_c$ are the complex frequency and coupling constant, which are expressed as $\tilde{\omega}_o = \omega_o + i(\Gamma_a + \Gamma_r)/2$ and $\tilde{\omega}_c = \alpha + i\beta$, where $\omega_o$ is the resonant angular frequency, $\Gamma_a$ and $\Gamma_r$ are the absorption and radiative decay rates, and α and β are the real and imaginary parts of the coupling constant. The inhomogeneous part $K^T\begin{bmatrix} s_{m,+} \end{bmatrix}$ defines the light from the continuum to drive $a_1$ and $a_2$ and their form depends on the number of available incident channels. For a given polarization, the reciprocal theorem requires the discrete incoming power amplitude vectors for the second and third gaps are $\begin{bmatrix} s_{m,+} \end{bmatrix} = \begin{bmatrix} s_{0,+} \end{bmatrix}$ and $\begin{bmatrix} s_{0,+} \\ s_{1,+} \end{bmatrix}$, respectively, following the earlier grating equation, as illustrated in Fig. 1(b) and (c). The second and third gap $K^T$ are given as $\begin{bmatrix} \tilde{\kappa}_{0,1} \\ \tilde{\kappa}_{0,2} \end{bmatrix}$ and $\begin{bmatrix} \tilde{\kappa}_{0,1} & \tilde{\kappa}_{1,1} \\ \tilde{\kappa}_{0,2} & \tilde{\kappa}_{1,2} \end{bmatrix}$, where the first and second subscripts of $\tilde{\kappa}_{m,n}$ are the diffraction order m and mode 1 or 2, and they define the complex in-coupling constant matrices.

We will diagonalize Eq. (1) to provide a clearer physical picture. After solving the homogeneous part of Eq. (1), the complex frequencies of the supermodes are $\tilde{\omega}_{+/-} = \omega_{+/-} + i\Gamma_{+/-} = (\omega_o \pm \alpha) + i((\Gamma_a + \Gamma_r)/2 \pm \beta)$, indicating their spectral positions and decay rates depend on $\alpha$ and $\beta$. We see the real part, $\omega_{+/-}$, of the supermodes are determined by the magnitude and sign of $\alpha$ and they are separated by a gap = $2\alpha$. On the other hand, for the imaginary part, $\Gamma_{+/-}$, one supermode has larger decay rate whereas another one has lower, featuring the bright and dark modes [43]. As the matrix is non-Hermitian and symmetric, we then solve for the left eigenvectors to find the transformation matrix T to be $\sqrt{\frac{1}{2}}\begin{bmatrix} 1 & 1 \\ 1 & -1 \end{bmatrix}$, and it transforms $a_{1/2}$ to the supermodes as $\begin{bmatrix} a_+ \\ a_- \end{bmatrix} = \sqrt{\frac{1}{2}} \begin{bmatrix} a_1 + a_2 \\ a_1 - a_2 \end{bmatrix}$, which are orthogonal despite the non-hermiticity [42]. Eq. (1) is then diagonalized as $\frac{d}{dt}\begin{bmatrix} a_+ \\ a_- \end{bmatrix} = i\begin{bmatrix} \tilde{\omega}_+ & 0 \\ 0 & \tilde{\omega}_- \end{bmatrix}\begin{bmatrix} a_+ \\ a_+ \end{bmatrix} + TK^T [s_{m,+}]$, indicating $a_{+/-}$ are now driven individually without any interaction between them. The Fourier space output channels can also be formulated properly. By using conservation of energy and time reversal symmetry, the outgoing radiation channels are expressed as $[s_{m,-}] = C[s_{m,+}] + K\begin{bmatrix} a_1 \\ a_2 \end{bmatrix} = C[s_{m,+}] + KT\begin{bmatrix} a_+ \\ a_- \end{bmatrix}$, where C is the non-resonant scattering matrix [41,42]. Apparently, the radiations from $a_{+/-}$ are manifested by KT, which results from the superposition between the radiations from the uncoupled $a_{1/2}$. Therefore, $[s_{m,-}]$ can be rewritten as:

$$s_{0,-} = \tilde{c} s_{0,+} + \sqrt{\frac{1}{2}}[\tilde{\kappa}_{0,1} + \tilde{\kappa}_{0,2}]a_+ + \sqrt{\frac{1}{2}}[\tilde{\kappa}_{0,1} - \tilde{\kappa}_{0,2}]a_- \qquad (2)$$

and

$$\begin{bmatrix} s_{0,-} \\ s_{1,-} \end{bmatrix} = \begin{bmatrix} \tilde{c}_1 & \tilde{c}_2 \\ \tilde{c}_2 & \tilde{c}_1 \end{bmatrix} \begin{bmatrix} s_{0,+} \\ s_{1,+} \end{bmatrix} + \sqrt{\frac{1}{2}} \begin{bmatrix} \tilde{\kappa}_{0,1} + \tilde{\kappa}_{0,2} \\ \tilde{\kappa}_{1,1} + \tilde{\kappa}_{1,2} \end{bmatrix} a_+ + \sqrt{\frac{1}{2}} \begin{bmatrix} \tilde{\kappa}_{0,1} - \tilde{\kappa}_{0,2} \\ \tilde{\kappa}_{1,1} - \tilde{\kappa}_{1,2} \end{bmatrix} a_-, \tag{3}$$

indicating each supermode has one single output channel at the second gap, but two at the third gap.

As the system possesses inversion symmetry, $a_{+/-}$ should carry odd and even spatial field symmetries with respect to the unit cell center provided they are orthogonal to each other [20]. Such feature eases the determination of γ. Without knowing the exact field symmetries of $a_{+/-}$, we expect the γ of a continuous band is 0 if both the circled supermodes at the Γ and X points in Fig. 1(a) are either $a_+$ or $a_-$ but is π if both $a_+$ and $a_-$ are found. To determine whether $a_+$ or $a_-$ is present, we further simplify Eq. (2) and (3) based on the relationships between $\tilde{\kappa}_{m,n}$ of a mirror symmetric diffraction pair. As an example, for the third gap in Fig. 1(c), the system symmetry requires $|\tilde{\kappa}_{0,1}| = |\tilde{\kappa}_{1,2}|$ and $|\tilde{\kappa}_{0,2}| = |\tilde{\kappa}_{1,1}|$ for $a_{1/2}$ propagating in the opposite directions. Given the polarizations of the radiations from $a_{1/2}$ scale with $\tilde{\kappa}_{m,n}$ and the pair from each mode should have the polarizations pointing either in the same clockwise or counterclockwise direction, as indicated in Fig. 1(c), to ensure continuity, the symmetry further requires two pairs carry opposite polarization directions, or $\tilde{\kappa}_{0,1} = -\tilde{\kappa}_{1,2}$ and $\tilde{\kappa}_{0,2} = -\tilde{\kappa}_{1,1}$. In fact, as we see in the Supplementary Information, for any mirror symmetric m = p and q pair, $\tilde{\kappa}_{p,1} = -\tilde{\kappa}_{q,2}$ and $\tilde{\kappa}_{p,2} = -\tilde{\kappa}_{q,1}$ [38]. As a result, Eq (3) can be simplified as

$$\begin{bmatrix} s_{0,-} \\ s_{1,-} \end{bmatrix} = \begin{bmatrix} \tilde{c}_1 & \tilde{c}_2 \\ \tilde{c}_2 & \tilde{c}_1 \end{bmatrix} \begin{bmatrix} s_{0,+} \\ s_{1,+} \end{bmatrix} + \sqrt{\frac{1}{2}} \begin{bmatrix} \tilde{\kappa}_{0,1} - \tilde{\kappa}_{1,1} \\ -(\tilde{\kappa}_{0,1} - \tilde{\kappa}_{1,1}) \end{bmatrix} a_+ + \sqrt{\frac{1}{2}} \begin{bmatrix} \tilde{\kappa}_{0,1} + \tilde{\kappa}_{1,1} \\ \tilde{\kappa}_{0,1} + \tilde{\kappa}_{1,1} \end{bmatrix} a_-. \tag{4}$$

Likewise, by applying the same argument to the second gap such that $\tilde{\kappa}_{0,1} = -\tilde{\kappa}_{0,2}$, we rewrite Eq. (2) as

$$s_{0,-} = \tilde{c}s_{0,+} + \sqrt{2}\tilde{\kappa}_{0,1}a_-, \tag{5}$$

where $a_+$ is always presented as a quasi-BIC whereas $a_-$ is a bright mode, making $a_+$ to be the symmetry protected BIC regardless of what the values of $\begin{bmatrix} \tilde{\kappa}_{0,1} & \tilde{\kappa}_{0,2} \end{bmatrix}^T$ are [32]. On the other hand, while the supermodes at the third gap in Eq. (4) are bright and dark, their assignment is difficult and relies on the interplay between the sign and magnitude of $\tilde{\kappa}_{0,1}$ and $\tilde{\kappa}_{1,1}$, which are strongly system geometry dependent. We see $a_+$ is the bright mode if $\tilde{\kappa}_{0,1}\tilde{\kappa}_{1,1} < 0$ but is dark when $> 0$. Interestingly, $a_+$ $(a_-)$ can also be quasi-BIC if $|\tilde{\kappa}_{0,1}| = |\tilde{\kappa}_{1,1}|$ $(|\tilde{\kappa}_{0,1}| = -|\tilde{\kappa}_{1,1}|)$, but such condition can only be met for certain geometry, which requires careful tuning of the system parameters.

We then explicitly formulate the discrete far-field channels. By considering only one single incident channel such that $\begin{bmatrix} s_{m,+} \end{bmatrix} = \begin{bmatrix} s_{0,+} \end{bmatrix}$ and $\begin{bmatrix} 0 \\ s_{1,+} \end{bmatrix}$, the scattering coefficients of the supermodes can be formulated as:

$$\frac{s_{0,-}}{s_{0,+}} = \tilde{c} + \frac{2\tilde{\kappa}_{0,1}}{i(\omega - \tilde{\omega}_-)} \tag{6}$$

for the second gap, and

$$\frac{s_{0,-}}{s_{1,+}} = \tilde{c}_2 - \frac{1}{2}\frac{(\tilde{\kappa}_{0,1} - \tilde{\kappa}_{1,1})^2}{i(\omega - \tilde{\omega}_+)} + \frac{1}{2}\frac{(\tilde{\kappa}_{0,1} + \tilde{\kappa}_{1,1})^2}{i(\omega - \tilde{\omega}_-)},$$

$$\frac{s_{1,-}}{s_{1,+}} = \tilde{c}_1 + \frac{1}{2}\frac{(\tilde{\kappa}_{0,1} - \tilde{\kappa}_{1,1})^2}{i(\omega - \tilde{\omega}_+)} + \frac{1}{2}\frac{(\tilde{\kappa}_{0,1} + \tilde{\kappa}_{1,1})^2}{i(\omega - \tilde{\omega}_-)}, \tag{7}$$

for two mirror symmetric pair at the third gap. One sees the radiation losses from $a_{+/-}$ are Lorentzian, rendering Fano spectral profiles in overall [44]. In addition, it is seen from Eq. (7) that the radiation pair from $a_+$ are $\pi$ out of phase whereas those from $a_-$ are in phase. Therefore, Eq. (6) and (7) provide good tool for determining the spectral positions of $a_{+/-}$ by measuring the angle- and wavelength-resolved complex diffraction mapping of the system in Fourier space. While the supermodes at the second gap can be visually inspected for determining the position of the quasi-BIC, i.e. $a_+$, the $\omega_{+/-}$ of the third gap are estimated by fitting the magnitude and phase, $|s_{0/1,-}/s_{1,+}|^2$ and $\arg(s_{0/1,-}/s_{1,+})$, spectra of the diffraction orders with Eq. (7). The extension of the CMT to any mirror symmetric pair at higher order band gap is provided in the Supplementary Information [38].

### III. FINITE-DIFFERENCE TIME DOMAIN SIMULATION

We verify the CMT model by FDTD simulations. Two types of optical systems are considered, and they are 1D Au plasmonic and $SiO_2$/Au photonic crystals. While the plasmonic crystals (PmCs) support TM-polarized Bloch-like SPPs [45], the photonic crystals (PhCs) excite TE waveguide modes [46]. We will present the results of PmCs here and those of the PhCs are provided in the Supplementary Information [38]. For the PmCs, the unit cell is shown in Fig. 2(a), with the period P and groove height H are set at 900 nm and 50 nm, respectively, and the groove width W is varied from 100 and 700 nm with a step size of 150 nm. The corresponding TM-polarized k- and wavelength-resolved total reflectivity, which sums all the diffraction orders, mappings are calculated along the Γ-X direction in Fig 2(b) – (f), showing the dispersive ±1 and -2 Bloch-like SPP bands follow the phase-matching equation given as $\frac{\varepsilon_{Au}}{\varepsilon_{Au}+1}\left(\frac{1}{\lambda}\right)^2 = \left(\frac{k}{2\pi}+\frac{n_B}{P}\right)^2$, where $\varepsilon_{Au}$ is the dielectric constant of Au, as illustrated by the dash lines in Fig 2(b) [33]. More importantly, one sees ±1 SPPs cross at the Γ point and +1 and -2 SPPs cross at the X point, forming two band gaps at λ = 925 and 650 nm and the supermodes. In agreement with the CMT model, the supermodes exhibit dark (high reflectivity) and bright (low reflectivity) radiation characteristics.

We will determine the Zak phase of the +1 SPP band, which is the band of interest in Fig. 1(a). At the Γ point for all PmCs, a quasi-BIC is always present, and it can be visually identified at the +1 band for W = 100 – 400 nm but flips to the -1 band when W increases further. The corresponding reflectivity spectra are plotted in Fig. 3(a), clearly showing only one single reflectivity dip is presented as the bright mode. As a result, we conclude $a_+$ locates at the +1 band for W = 100 – 400 nm. On the other hand, at the X point, we no longer can differentiate the spectral positions of $a_{+/-}$ simply by examining the reflectivity spectra. We simulate the diffraction orders and find only the m = 0 and 1 mirror symmetric orders exist in the continuum, following the grating equation well. The corresponding $|s_{0/1,-}/s_{1,+}|^2$ and $\arg(s_{0/1,-}/s_{1,+})$ spectra are calculated in Fig. 3(b) and (c) and fitted with Eq. (7) to determine the $\omega_{+/-}$, $\Gamma_{+/-}$, $\tilde{\kappa}_{0,1} - \tilde{\kappa}_{1,1}$, and $\tilde{\kappa}_{0,1} + \tilde{\kappa}_{1,1}$ of two supermodes by assuming both the $\tilde{c}_1(\lambda)$ and $\tilde{c}_2(\lambda)$ backgrounds are polynomial functions. The best fits are displayed as the solid lines. All fitted $\omega_{+/-}$ of the PmCs are summarized in Table 1, where the high and low energy modes at two high symmetry points that correspond to the modes sitting on the +1 band are highlighted. Reminding if two highlights are either $a_+$ or $a_-$, the Zak phase is 0, but π when they are different [17,20]. We see γ = π for W = 100, 250, 550 nm but γ = 0 for 400 and 700 nm.

To confirm our findings, we have simulated the $|E_z|$ near-field intensity profiles at the zone center and boundary of the +1 band by FDTD in Fig. 4(a) and (b) for different W. At the zone center, we see the $|E_z|$ profiles are even with respect to the groove center for W = 100 – 400 nm but change to odd afterwards. On the other hand, the profiles at the zone boundary are odd for W = 100, 250, and 700 nm but are even for 400 and 550 nm. We see despite the systems are non-Hermitian, the field symmetries of the supermodes are either odd or even [20]. As a result, the comparison between the field symmetries at the zone center and boundary indicates γ = π for W = 100, 250 and 550 nm but 0 for 400 and 700 nm, in consistent with the far-field determination.

## IV. EXPERIMENTAL VERIFICATION

A series of 1D periodic Au rectangular groove PmCs has been fabricated by focused ion beam (FIB) and their scanning electron microscopy (SEM) images are shown in the insets of Fig. 5(a) – (e), showing they have P = 900 nm, H = 50 nm, and W varying from 100 to 700 nm. After the sample preparation, the PmCs are then transferred to a homebuilt Fourier space optical microscope described in the Supplementary Information for angle- and wavelength-resolved diffraction measurements [38]. Briefly, a supercontinuum generation laser is illuminated on the sample at a well-defined incident angle θ via the microscope objective lens and the signals from the sample are collected by the same objective lens in which the diffraction orders are projected onto the Fourier space [47,48]. By using an aperture to filter out the desired diffraction order, a spectrometer-based CCD detector and a common path interferometer are used for measuring the magnitude and phase spectra [49,50].

By varying θ sequentially and at the same time measuring the total reflection spectra, we contour plot the TM-polarized reflectivity mappings in Fig. 5(a) – (e) for different W along the Γ-X direction. They show ±1 and -2 SPP bands are present, and the bands are consistent with the phase-matching equation as illustrated by the dash lines. From the mappings, we see at normal incidence, or the Γ point, BIC-like mode is always observed near the band gap. The +1 band has $a_+$ for W = 100 – 400 nm but $a_-$ for larger W after visual examination. On the other hand, at the X point where +1 and -2 SPPs cross at $\theta_i$ ~ 20.5º, we see the dark and bright modes are found and their positions depend on W. To estimate the spectral positions of $a_{+/-}$, we measure the corresponding m = 0 and 1 diffraction and TM-TE phase difference spectra in Fig. 6(a) and (b) and fit them by using Eq. (7) to determine the $\omega_{+/-}$ in Table 1, which shows the +1 band is $a_-$ for W = 100, 250 and 700 nm is $a_+$ for 400 and 550 nm. Therefore, γ = π for W = 100, 250 and 550 nm but = 0 for 400 and 700 nm.

Finally, we demonstrate a topologically protected state is formed at the interface between two topological trivial and nontrivial PmCs [6,18]. We construct a heterostructure by joining two W = 100 and 400 nm PmCs together. In prior to joining, we have examined by FDTD the field symmetries at the zone center and boundary of two PmCs and determine the $\gamma_\ell$ of the $\ell$ = 0, -1, and +1 SPP bands to be $\pi$, $\pi$ and $\pi$ for W = 100 nm and $\pi$, $\pi$ and 0 for W = 400 nm. Therefore, the sums of $\gamma_\ell$ give $3\pi$ and $2\pi$ for W = 100 and 400 nm PmCs, indicating the -2/+1 energy gaps at the zone boundary are topological nontrivial and trivial. We then simulate the heterostructure supercell as shown in Fig. 7(a) that consists of 14 unit cells of W = 100 and 400 nm PmCs on the right- and left-handed sides [51]. Fig. 7(b) shows the TM-polarized k- and wavelength-resolved reflectivity mapping at the zone boundary along the Γ-X direction, clearly demonstrating a localized mode is located at k = 0.5 or $\theta_i$ = 20.5° and λ ~ 640 nm in the mid of the band gap. We also have simulated the wavelength-dependent near-field mapping of the heterostructure. For different wavelengths, the near-field intensities at 20 nm above the surface is simulated across the heterostructure and then contour plotted in Fig. 7(c), showing the interface is located at x = 0 μm and the trivial and nontrivial regions are at x > 0 and < 0 μm, respectively. One sees two strong fields are visible at ~ 620 and 670 nm in the PmC bulk regions away from the interface due to the excitations of the upper and lower coupled modes. However, the strongest field strength is observed at the interface, x = 0 μm, at 640 nm, and it decays rapidly into the bulk regions, signifying the presence of a topologically protected interface state [51]. We have prepared the heterostructure by FIB and its SEM image is shown in Fig. 7(d) with W = 100 and 400 nm PmCs on the right- and left-hand sides. The TM-polarized k- and wavelength-resolved reflectivity mapping of the sample is illustrated in Fig. 7(e), showing a localized state is found at $\theta_i$ = 20.5° and λ ~ 625 nm in the +1/-2 band gap at the zone boundary. We expect it is an interface state as projected from the simulation.

## V. CONCLUSION

In summary, we have developed a temporal CMT model to determine the Zak phase of an isolated band in 1D leaky photonic systems. At the Brillouin zone center and boundary, we find the radiation losses from the supermodes in Fourier space present different characteristics ranging from quasi-BIC to odd or even mirror symmetric diffraction pair, which are strongly associated with the spatial field symmetries and thereby the corresponding Zak phase. For verification, 1D PmCs and PhCs that support TM- and TE-polarized SPP and waveguide modes have been studied by FDTD and the results agree very well with the theory. We also have prepared 1D PmCs by FIB and examined their diffractions by using Fourier space diffraction spectroscopy and common path interferometry for determining the Zak phases. In the end, a topological protected interface state is demonstrated by joining two topological trivial and nontrivial PmCs together.

## VI. ACKNOWLEDGMENT

This research was supported by the Chinese University of Hong Kong through Area of Excellence (AoE/P-02/12) and Innovative Technology Fund Guangdong-Hong Kong Technology Cooperation Funding Scheme (GHP/077/20GD) and Partnership Research Program (PRP/048/22FX).

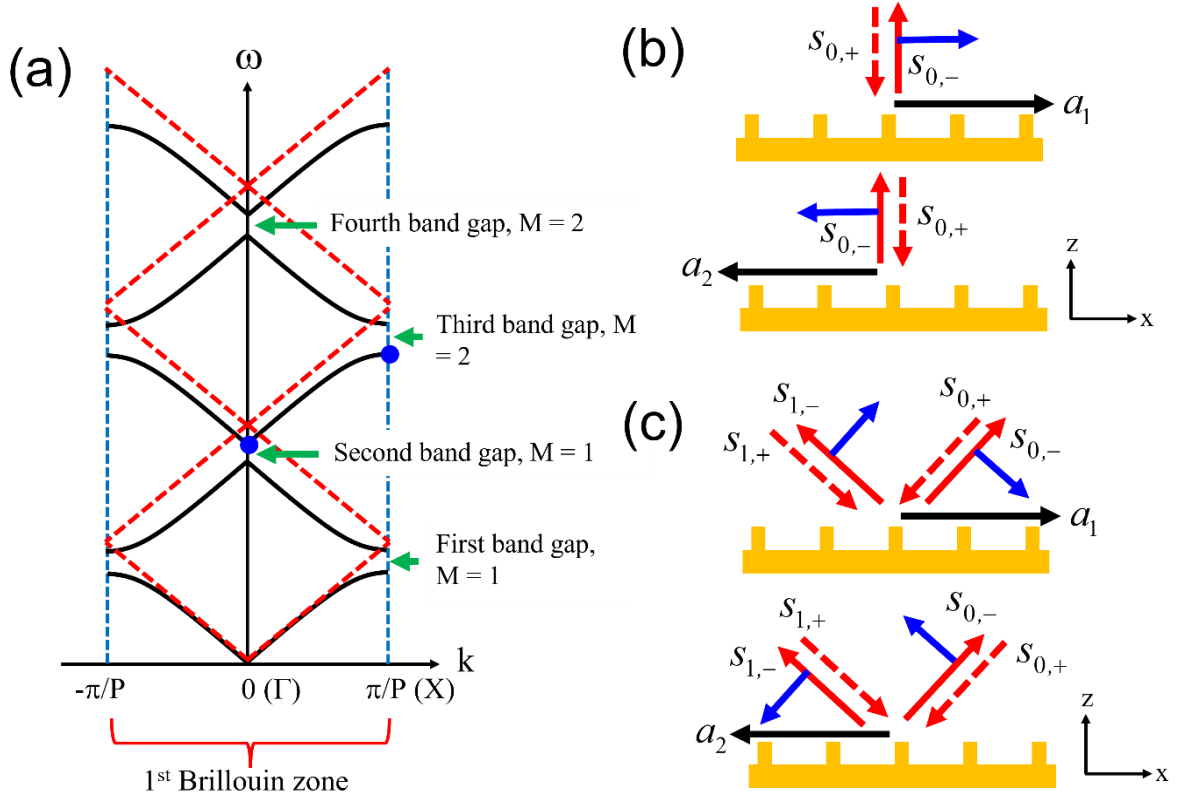

Fig. 1. (a) The ω-k plot of a 1D optical system that supports guided resonances. The solid and dashed lines are the resonant Bloch modes and the Wood's anomalies deduced by using the phase-matching equation. The Bloch modes form continuous energy bands that are split by band gaps at the Γ and X points in the first Brillouin zone. At the gap, two supermodes are formed at lower and higher energies. The Wood's anomalies cross at the Γ and X points following $\lambda_o = P/M$ and $P/(M-1/2)$. The supermodes of interest at the zone center and boundary are marked by blue solid circles. (b) At the second band gap, two Bloch-like modes $a_{1/2}$ propagate in the opposite x-directions, and each supports one m = 0 input and output ports $s_{0,\pm}$. The output radiation channels from $a_{1/2}$ carry linear polarizations pointing in opposite directions. (c) At the third gap, $a_{1/2}$ supports two mirror symmetric m = 0 and 1 input and output port $s_{0,\pm}$ and $s_{1,\pm}$ at $\pm\theta_i$. The polarizations of two output channels from $a_1$ point in the same clockwise and counterclockwise direction to ensure continuity but are opposite to those from $a_2$.

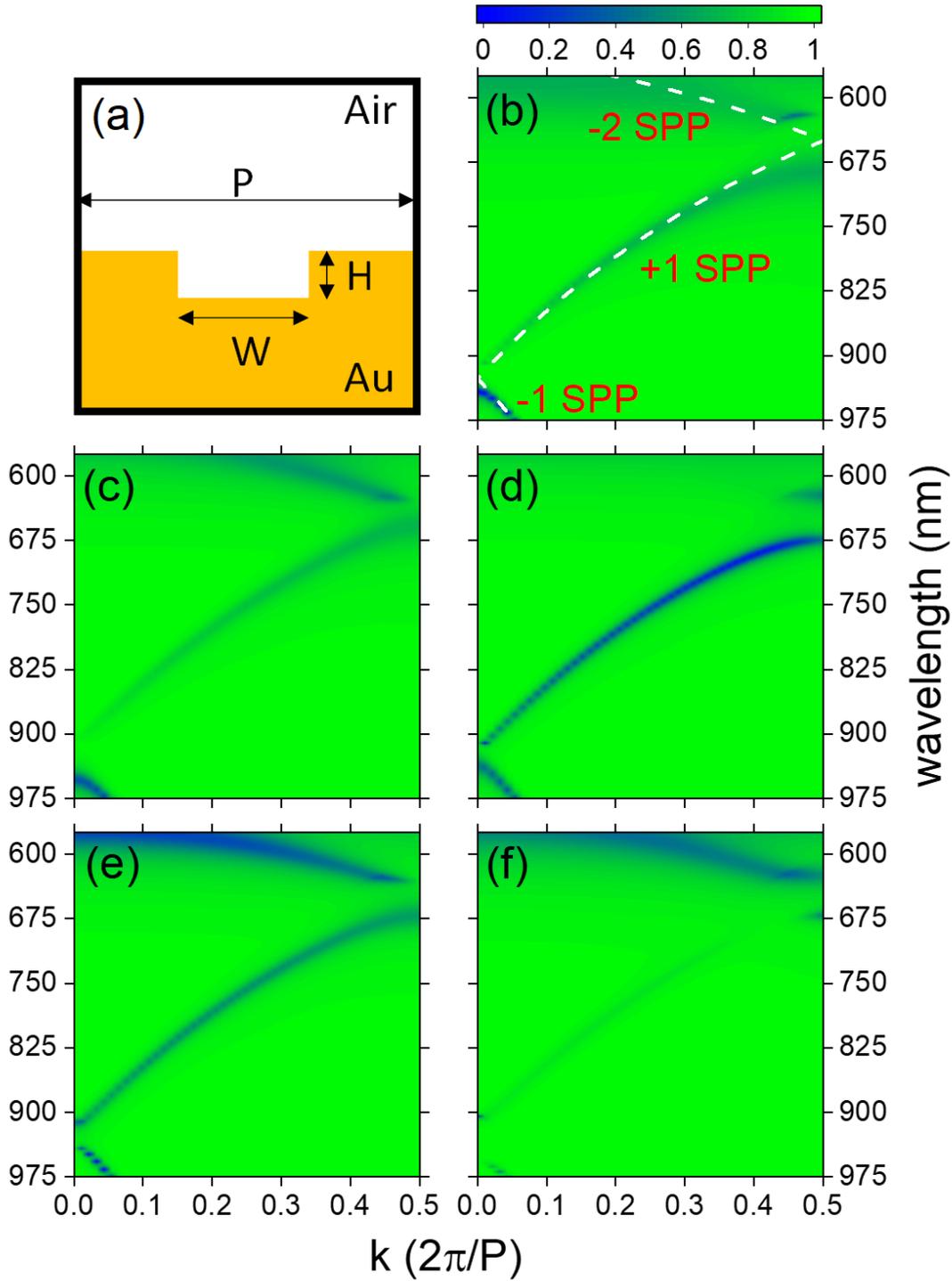

Fig. 2. (a) The unit cell of 1D PmC for FDTD simulations. The simulated TM-polarized k- and wavelength-resolved total reflectivity mappings of the PmCs with W = (b) 100, (c) 250, (d) 400, (e) 550, and (f) 700 nm taken along the Γ-X direction. The white dashed lines are calculated by using the phase-matching equation, indicating ±1 and -2 Bloch-like SPPs are excited. At the Γ and X points where k = 0 and 0.5, two energy band gaps are formed, featuring two dark and bright modes are located above or below the gap. Particularly, at k = 0, a quasi-BIC is observed at either above or below the gap.

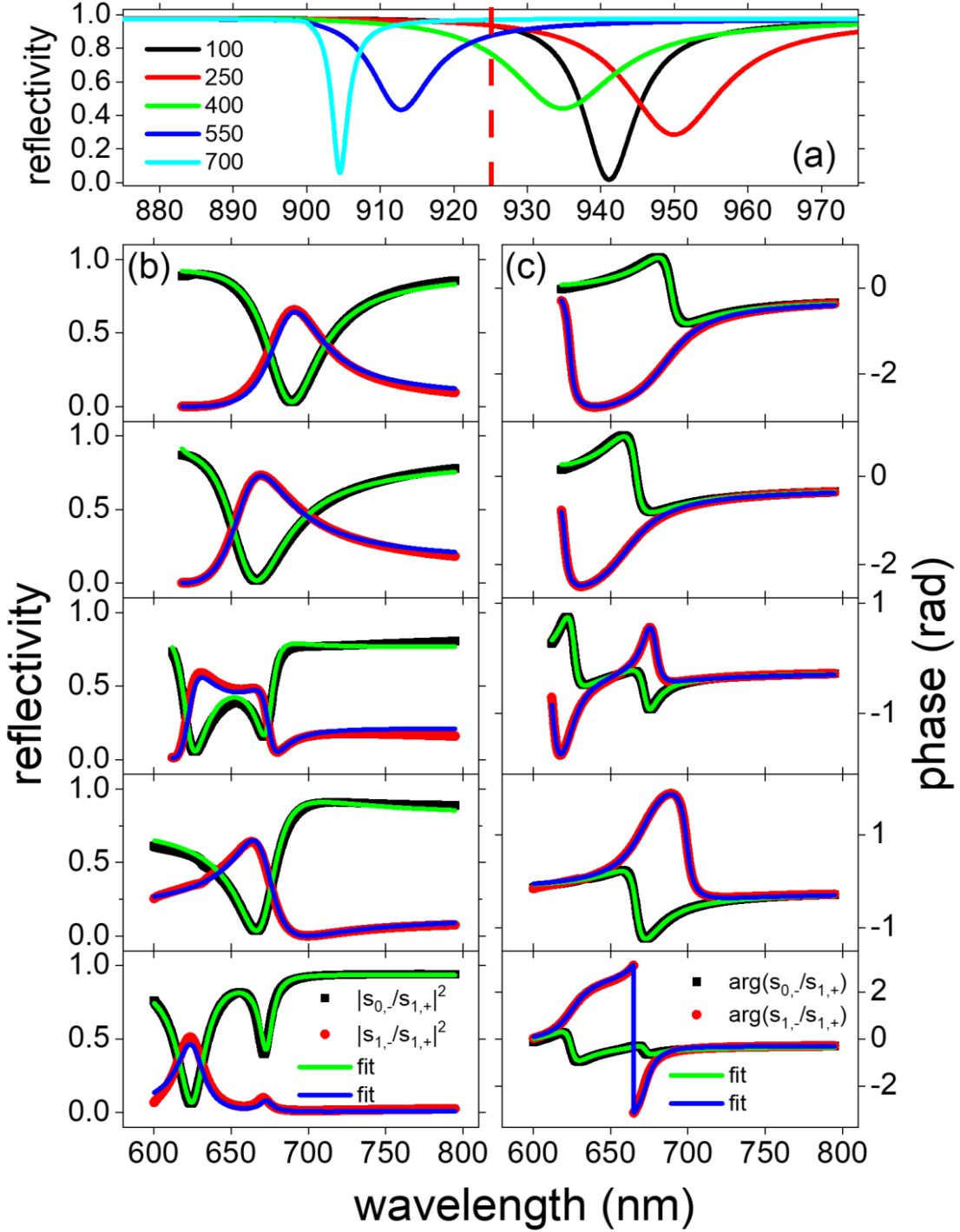

Fig. 3. The TM-polarized total reflectivity spectra of PmCs taken at the Γ point for different W, exhibiting only one single reflectivity dip as the bright mode. The red dash line is the band gap center, indicating the quasi-BIC occurs at shorter wavelength for W = 100, 250 and 400 nm but longer wavelength for W = 550 and 700 nm. At the X point, two TM-polarized mirror symmetric m = 0 (black square) and 1 (red circle) (b) reflectivity and (c) phase spectra for W = 100 (top) to 700 (bottom). The green and blue solid lines are the best fits determined by CMT.

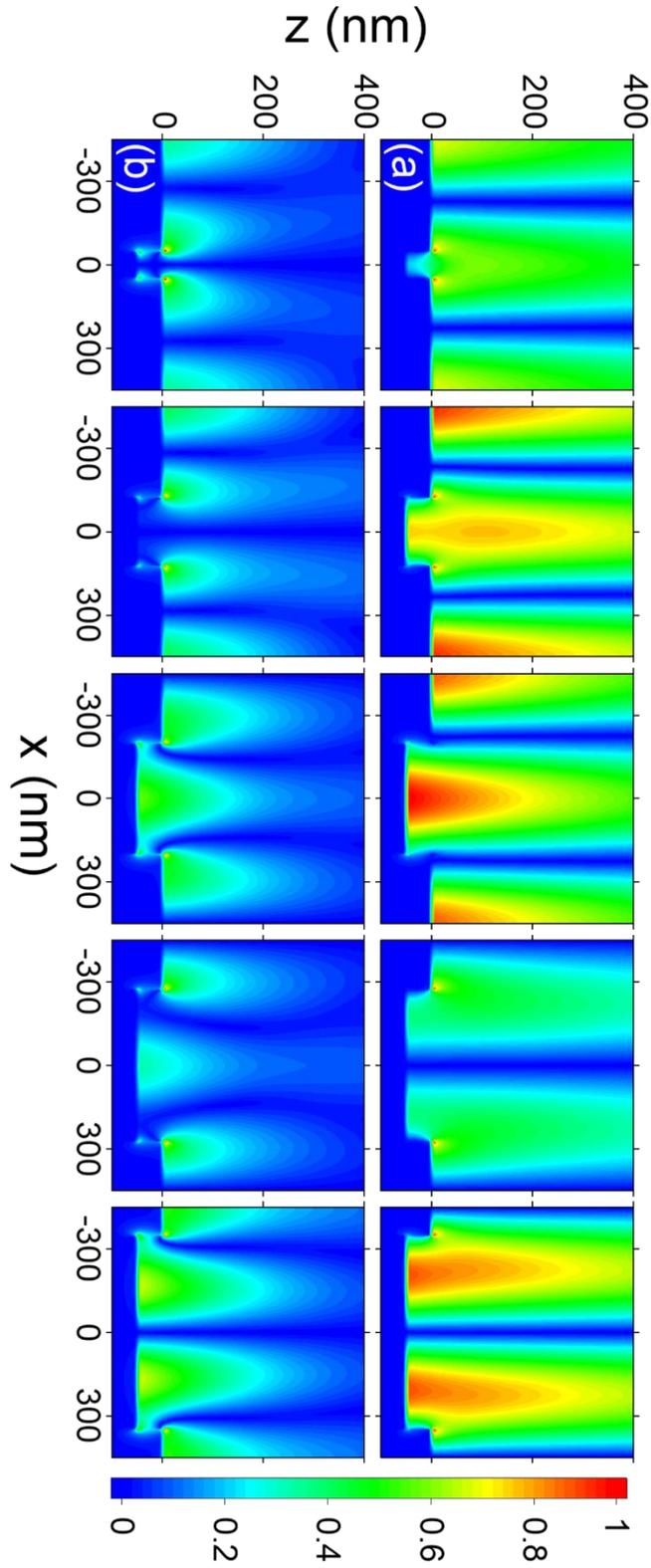

Fig. 4. The FDTD simulated $|E_z|$ near-field patterns of the PmCs for different W taken at the (a) Γ and (b) X points, showing their field symmetries are the same for W = 400 and 500 nm but different for W = 100, 250, and 700 nm.

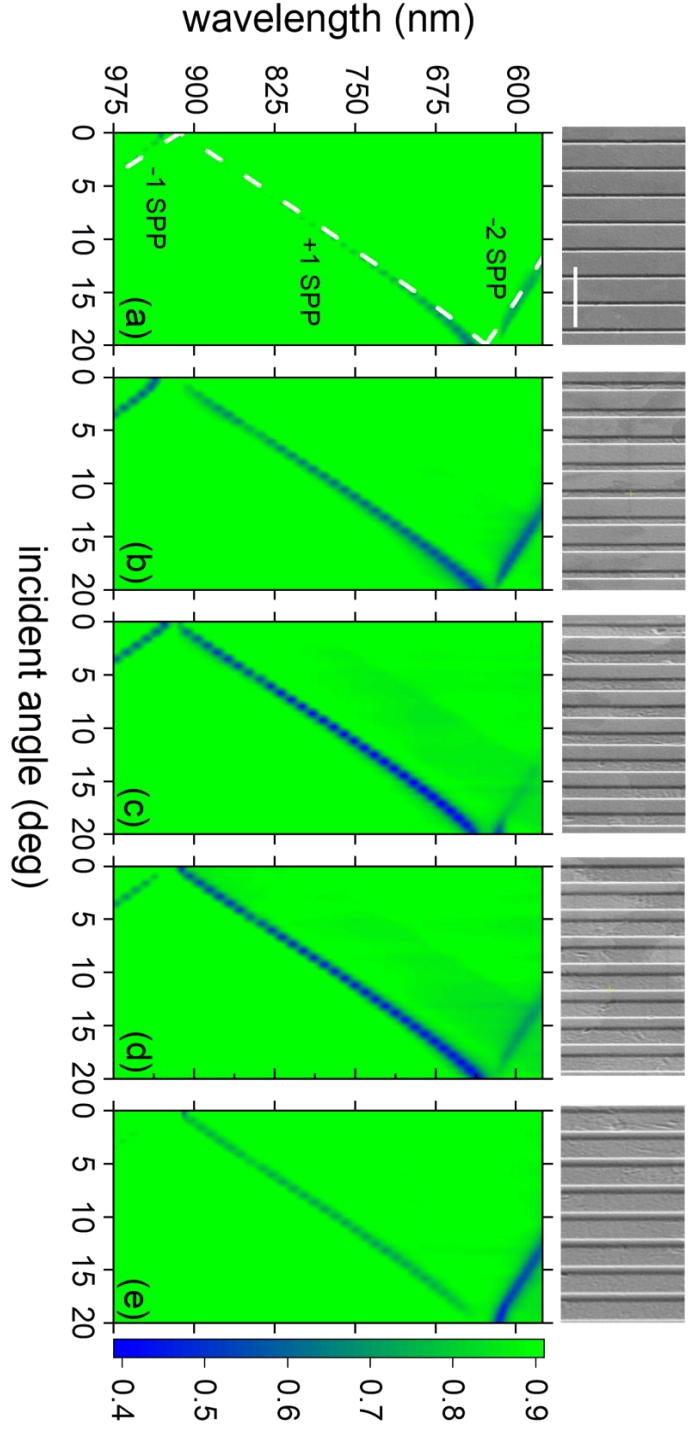

Fig. 5. The measured TM-polarized angle- and wavelength-resolved total reflectivity mappings of the PmCs with W = (a) 100, (b) 250, (c) 400, (d) 550, and (e) 700 nm taken along the Γ-X direction. The white dashed lines are ±1 and -2 Bloch-like SPPs determined by the phase matching equation. Two band gaps are formed at $\theta_i = 0°$ and ~ 20°, which correspond to the Γ and X points. The insets are the corresponding SEM images of the PmCs with the scale bare = 2 µm.

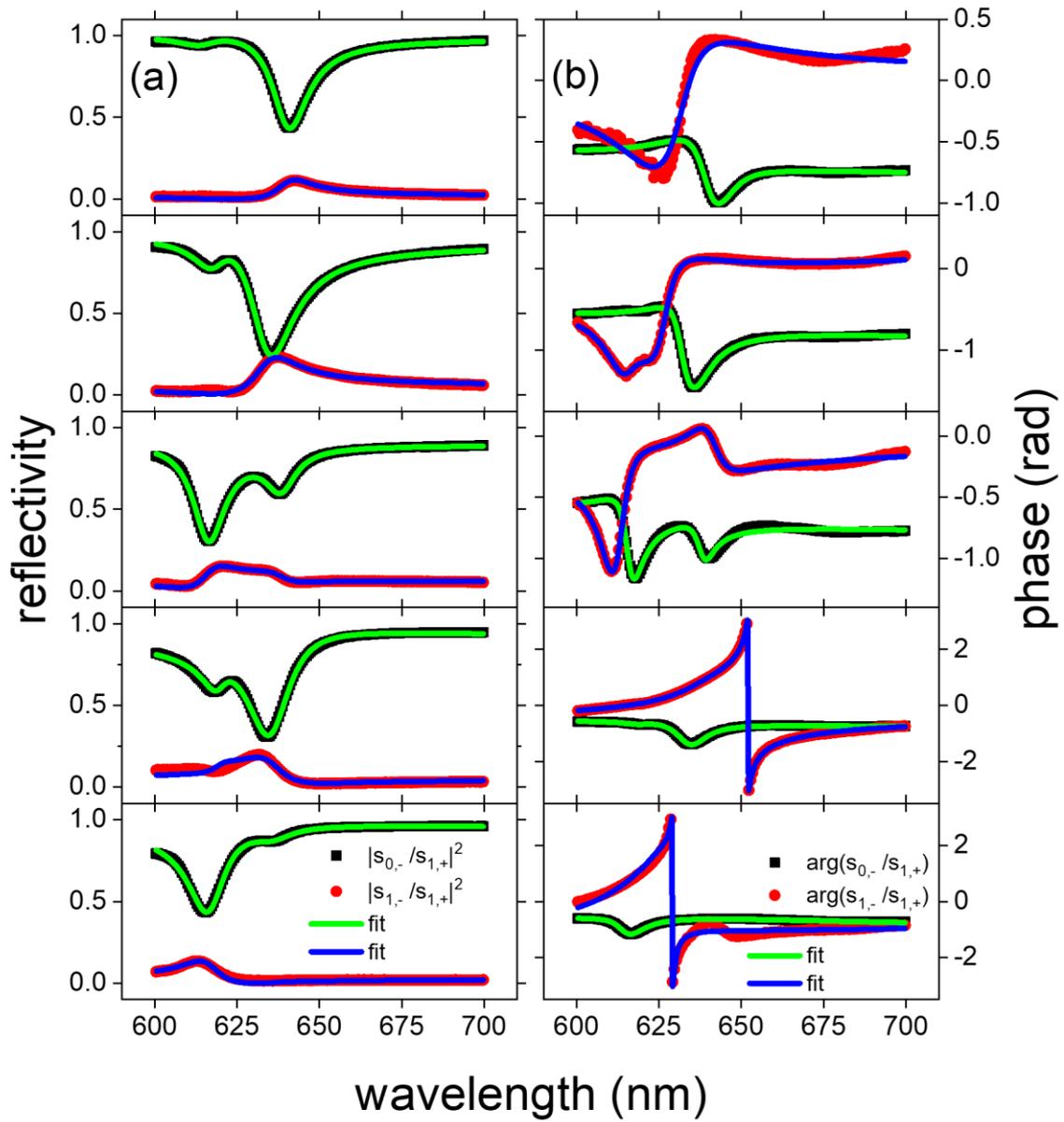

Fig. 6. At the zone boundary, two measured TM-polarized mirror symmetric m = 0 (black square) and 1 (red circle) (b) reflectivity and (c) TM-TE phase difference spectra for W = 100 (top) to 700 (bottom). The green and blue solid lines are the best fits determined by CMT.

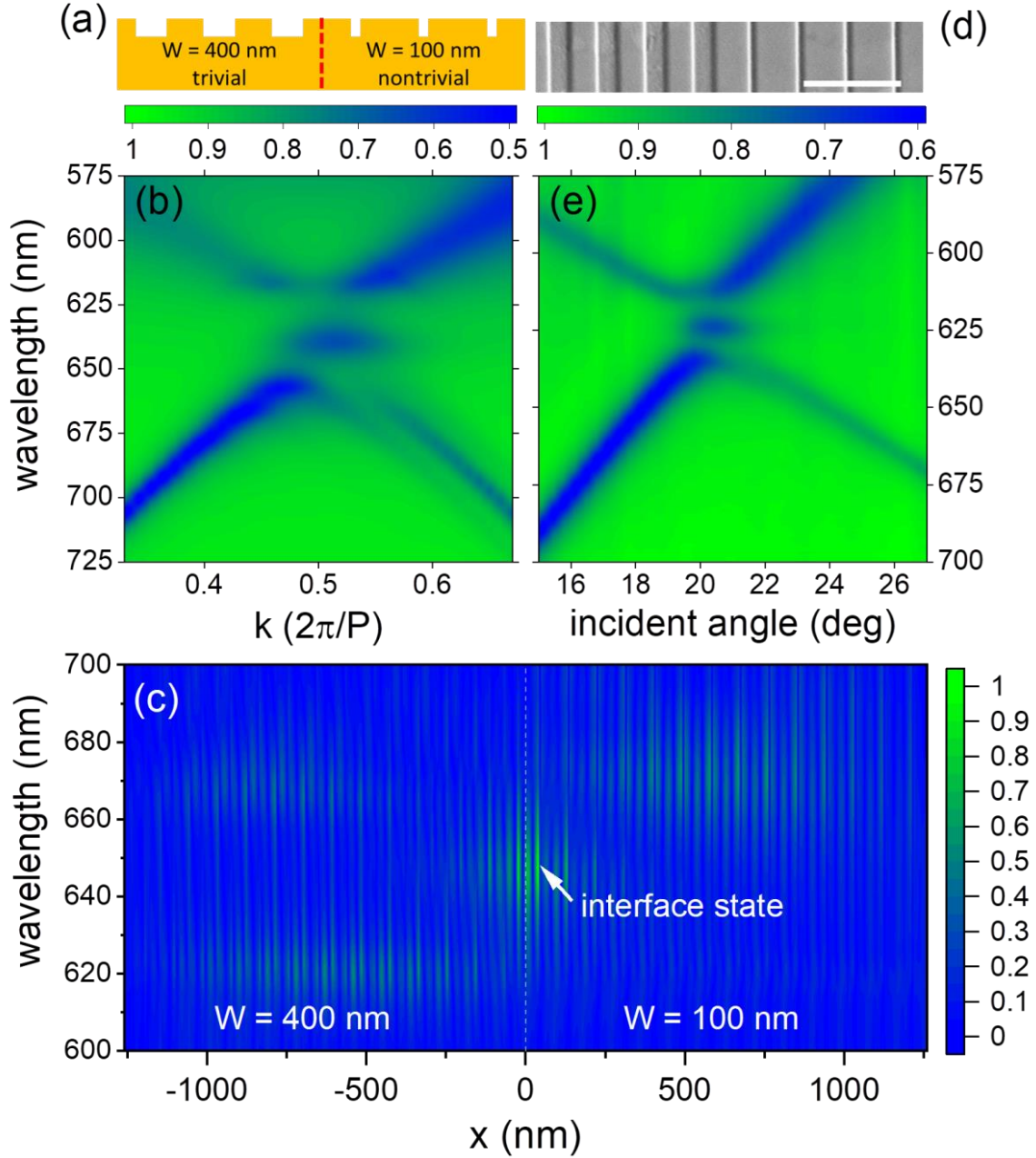

Fig. 7. (a) The schematic of the heterostructure by joining W = nontrivial 100 and trivial 400 nm PmCs. The interface is marked by the dash line. (b) The FDTD simulated TM-polarized reflectivity mapping of the heterostructure taken at the zone boundary along the Γ-X direction, showing an interface state is found within the gap at λ = 640 nm. (c) The wavelength-dependent near-field intensity mapping simulated at 20 nm above the heterostructure. The interface is located at x = 0 μm, showing strong field localization. The strong fields at 620 and 670 nm arise from the PmC bulk regions. (d) The SEM image of the W = 100 and 400 nm with the scale bar corresponding to 2 μm. (e) The measured TM-polarized angle-dependent reflectivity mapping of the heterostructure taken at the zone boundary along the Γ-X direction, showing an interface state is found within the gap at λ = 625 nm.

|  |  |  | 100 nm | 250 nm | 400 nm | 550 nm | 700 nm |
|---|---|---|---|---|---|---|---|
| FDTD | Γ, zone center | $\omega_+$ (eV) | 1.36 | 1.37 | 1.36 | 1.32 | 1.30 |
|  |  | $\omega_-$ (eV) | 1.32 | 1.31 | 1.33 | 1.36 | 1.37 |
|  | X, zone boundary | $\omega_+$ (eV) | 2.00 | 1.98 | 1.84 | 1.85 | 1.98 |
|  |  | $\omega_-$ (eV) | 1.82 | 1.89 | 1.99 | 1.96 | 1.85 |
| Experiment | Γ, zone center | $\omega_+$ (eV) | 1.36 | 1.36 | 1.36 | 1.33 | 1.32 |
|  |  | $\omega_-$ (eV) | 1.33 | 1.32 | 1.34 | 1.36 | 1.36 |
|  | X, zone boundary | $\omega_+$ (eV) | 2.02 | 2.01 | 1.95 | 1.96 | 2.02 |
|  |  | $\omega_-$ (eV) | 1.93 | 1.96 | 2.02 | 2.01 | 1.95 |

Table 1. The FDTD and experimental $\omega_\pm$ at the Γ and X points for the PmCs with different W. The highlights are the coupled modes located on the +1 SPP band. If the highlights at the zone center and boundary are both $a_+$ or $a_-$, the Zak phase is 0. If not, the Zak phase is π.

# Supplementary Information

# Determination of the Zak phase of one-dimensional photonic systems via far-field radiations in Fourier space

C. Liu, H.R. Wang, and H.C. Ong

Department of Physics, The Chinese University of Hong Kong, Shatin, Hong Kong, People's Republic of China

A. **Derivation of the number of radiation output channels by grating equation**

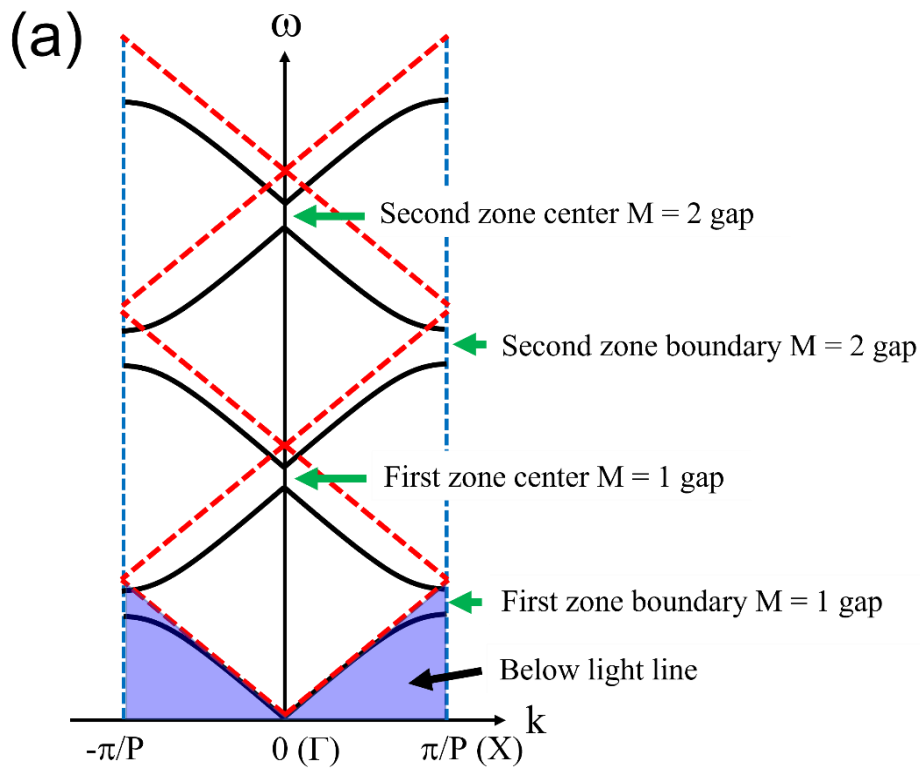

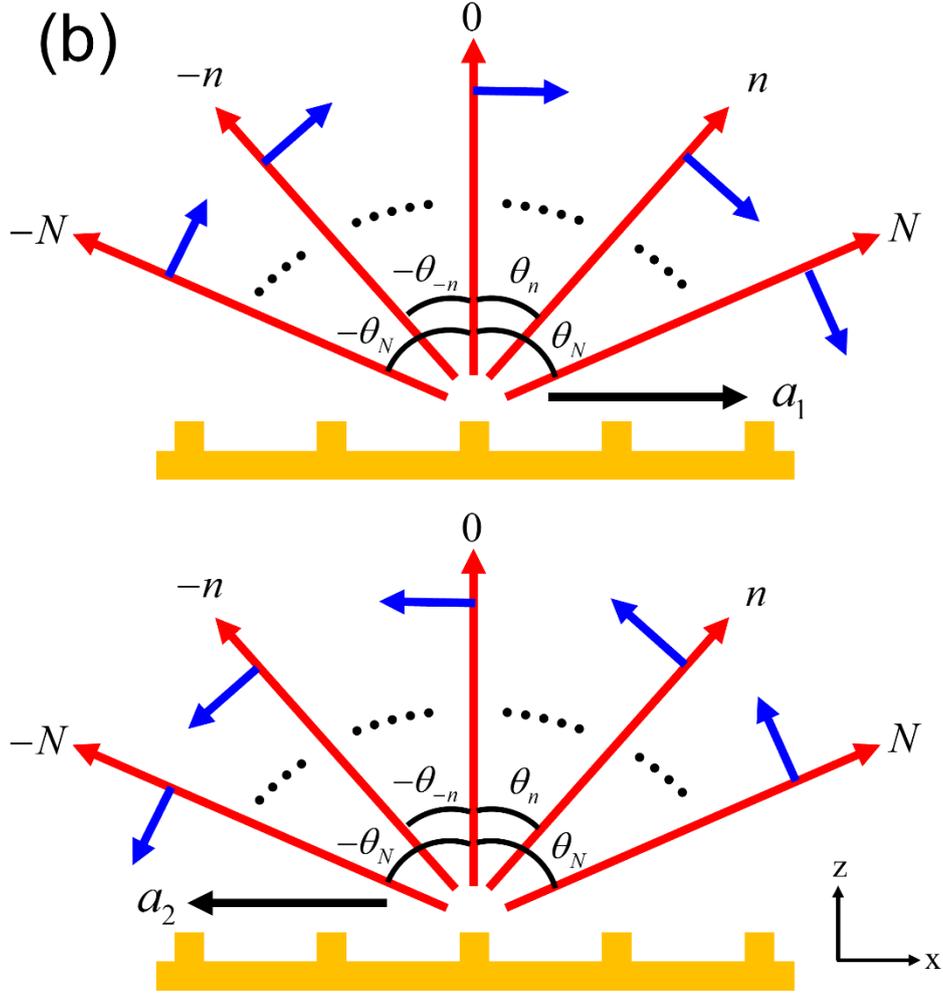

Fig. S1. (a) The ω-k plot of 1D optical system that support guided resonances. The solid and dashed lines are the resonant Bloch modes and the Wood's anomalies deduced by using phase-matching equation. The Bloch modes form continuous energy bands and are split by band gaps at the Brillouin zone center and boundary at k = 0 and $\pi/P$ μm$^{-1}$ to form two supermodes. The Wood's anomalies cross at the zone center and boundary following $\lambda_o = P/M$ and $P/(M-1/2)$. (b) Two Bloch-like modes $a_{1/2}$ propagate in opposite directions, and each mode supports $2N+1$ or $2N$ radiation channels, depending on whether it is located at the zone center or boundary. The normal channel is present (absent) at the zone center (boundary). However, there are always N pairs of mirror symmetric channels. The polarizations of all 2N channels from $a_1$ point in either the same clockwise or counterclockwise direction to ensure continuity, but they are opposite to those from $a_2$. At the Γ (X) point, the normal, i.e. 0, port is present (absent).

Fig. S1(a) shows the TM- or TE-polarized ω-k plot of a 1D periodic system that supports guided resonances along the Γ-X direction. The in-plane wavevector k is $2\pi \sin\theta_i/\lambda$, where $\theta_i$ is the incident angle. The optically thick system possesses inversion symmetry in the x-direction and is excited by light incident from the reflection side. The plot consists of the resonant Bloch modes and the Wood's anomalies as indicated by the solid and dashed lines and they are calculated by using the phase-matching equation given as $\left(n_{eff}\frac{\omega}{c}\right)^2 = \left(k + n_B\frac{2\pi}{P}\right)^2$, where $n_B$ is the Bragg scattering order, c is the light speed, and P and $n_{eff}$ are the period and the effective refractive index of the system, in analogous to the band folding scheme in electronic solid [1]. The $n_{eff}$ of the Wood's anomalies is taken to be 1 (air) [2]. We see at the Brillouin zone center ($\Gamma$ point) where k = 0 µm$^{-1}$, the Wood's anomalies cross at $\lambda_o = \frac{P}{M}$, where M is a nonzero positive integer that increases sequentially with the cross-point position. On the other hand, at the zone boundary (X point) where k = $\pi$/P µm$^{-1}$, the Wood's anomalies cross at $\lambda_o = \frac{P}{M-1/2}$. The Bloch modes cross at the $\Gamma$ and X points, yielding an energy band gap and two supermodes split at lower and higher energies. It is noted that the spectral positions of the supermodes are lower than the cross-points of Wood's anomalies because the $n_{eff}$ of the system is larger than air, i.e. $\lambda > \lambda_o$.

Once the supermode is excited, it will then dissipate radiatively via diffraction into discrete channels in Fourier space. For a reflective corrugated system, the grating equation is $m\lambda = P(\sin\theta_i + \sin\theta_m)$, where m is the diffraction order and $\theta_m$ is the diffraction angle [3]. At the $\Gamma$ point where $\theta_i$ = 0°, as shown in Fig. S1(a), the supermode has $\lambda > \lambda_o = \frac{P}{M}$, and the grating equation becomes $\frac{m}{M}\frac{\lambda}{\lambda_o} = \sin\theta_m$. It is then straightforward to show there are 2(M–1)+1 or 2N+1 propagating orders in the continuum. For example, for M = 1, i.e. the first gap, $m\frac{\lambda}{\lambda_o} = \sin\theta_m$ and there is only one m = 0 normal diffraction at $\theta_{m=0}$ = 0° available for each supermode. When M = 2 such that $\frac{m}{2}\frac{\lambda}{\lambda_o} = \sin\theta_m$ for the second gap, each supermode supports three diffraction orders, i.e. m = 0 and ±1, with $\theta_{m=0,\pm1}$ = 0° and $\pm\sin^{-1}\left(\frac{1}{2}\frac{\lambda}{\lambda_o}\right)°$. For the M = 3 third gap, five diffraction orders are present at $\theta_{m=0,\pm1,\pm2} = 0^o, \pm sin^{-1}\left(\frac{1}{3}\frac{\lambda}{\lambda_o}\right)^o$, and $\pm sin^{-1}\left(\frac{2}{3}\frac{\lambda}{\lambda_o}\right)^o$. As a result, for any M gap supermode at the zone center, one order is always the normal diffraction whereas the other N pairs are mirror symmetric with respect to the system surface normal, as shown in Fig. S1(b).

On the other hand, at the X point where $k = \frac{2\pi}{\lambda}sin\theta_i = \frac{\pi}{P}$, as shown in Fig. S1(a), the supermode of all energy gaps occurs at $\lambda > \lambda_o = \frac{P}{M - 1/2}$. The grating equation becomes $\frac{2m-1}{2M-1}\frac{\lambda}{\lambda_o} = sin\theta_m$, and we can show there are 2(M–1) or 2N diffraction orders available. For example, the first M = 1 gap, we have $(2m-1)\frac{\lambda}{\lambda_o} = sin\theta_m$ and no propagating diffraction is found in the continuum because the supermode is below the light line (Fig. S1(a)). For the M = 2 second gap, we have $\frac{2m-1}{3}\frac{\lambda}{\lambda_o} = sin\theta_m$ and each supermode supports two m = 0 and 1 diffraction orders with $\theta_{m=0,1} = \pm sin^{-1}\left(\frac{1}{3}\frac{\lambda}{\lambda_o}\right)^o$. For the M = 3 third gap, we have $\frac{2m-1}{5}\frac{\lambda}{\lambda_o} = sin\theta_m$ and four m = 0, ±1 and 2 are the possible solutions with $\theta_{m=0,\pm1,2} = \pm sin^{-1}\left(\frac{1}{5}\frac{\lambda}{\lambda_o}\right)^o$ and $\pm sin^{-1}\left(\frac{2}{5}\frac{\lambda}{\lambda_o}\right)^o$. Therefore, for any M gap supermode at the zone boundary, there are N mirror symmetric diffraction pairs. In contrast to the diffraction orders at the zone center, the normal diffraction order is always missing here.

B. **Generalization of the radiation channels of the supermode at the Brillouin zone center, Γ point.**

We first solve the radiation outputs of the supermode at the Γ point. The dynamics of two counter-propagating Bloch mode amplitudes, $a_1$ and $a_2$, taken under TM or TE polarization can be written as:

$$\frac{d}{dt}\begin{bmatrix} a_1 \\ a_2 \end{bmatrix} = i \begin{bmatrix} \tilde{\omega}_o & \tilde{\omega}_c \\ \tilde{\omega}_c & \tilde{\omega}_o \end{bmatrix} \begin{bmatrix} a_1 \\ a_2 \end{bmatrix} + K^T [s_+], \tag{S1}$$

where $\tilde{\omega}_o$ and $\tilde{\omega}_c$ are the complex frequency and coupling constant, which are expressed as $\tilde{\omega}_o = \omega_o + i(\Gamma_a + \Gamma_r)/2$ and $\tilde{\omega}_c = \alpha + i\beta$, where $\omega_o$ is the resonant angular frequency, $\Gamma_a$ and $\Gamma_r$ are the absorption and radiative decay rates, and $\alpha$ and $\beta$ are the real and imaginary parts of the coupling constant. For a given polarization, as aforementioned in the previous section, for any M gap, there are 2N+1 propagating diffraction orders at the $\Gamma$ point. The discrete incoming power amplitude vector thus can be expressed as $[s_+] = [s_{-N,+} \quad s_{-N+1,+} \quad \cdots \quad s_{0,+} \quad \cdots \quad s_{N-1,+} \quad s_{N,+}]^T$, where $s_{0,+}$ is denoted as the surface normal power and $s_{\pm n,+}$ are the mirror symmetric pair powers in which n is an integer $\leq$ N [4]. The complex in-coupling constant matrices is then written as

$$K_\Gamma^T = \begin{bmatrix} \tilde{\kappa}_{-N,1} & \tilde{\kappa}_{-N+1,1} & \cdots & \tilde{\kappa}_{0,1} & \cdots & \tilde{\kappa}_{N-1,1} & \tilde{\kappa}_{N,1} \\ \tilde{\kappa}_{-N,2} & \tilde{\kappa}_{-N+1,2} & \cdots & \tilde{\kappa}_{0,2} & \cdots & \tilde{\kappa}_{N-1,2} & \tilde{\kappa}_{N,2} \end{bmatrix},$$

where the subscript $\Gamma$ stands for the $\Gamma$ point and the first and second subscripts of $\tilde{\kappa}$ are the n channel and the mode 1 or 2, for inputting energy from the continuum to $a_1$ and $a_2$.

For non-Hermitian and symmetric matrix, we solve the left eigenvectors and the transformation matrix to be $T = \sqrt{\frac{1}{2}}\begin{bmatrix} 1 & 1 \\ 1 & -1 \end{bmatrix}$. Eq. (S1) can then be diagonalized as:

$$\frac{d}{dt}\begin{bmatrix} a_+ \\ a_- \end{bmatrix} = i \begin{bmatrix} \tilde{\omega}_+ & 0 \\ 0 & \tilde{\omega}_- \end{bmatrix} \begin{bmatrix} a_+ \\ a_- \end{bmatrix} + TK_\Gamma^T [s_+], \tag{S2}$$

where

$$TK_\Gamma^T = \sqrt{\frac{1}{2}} \begin{bmatrix} \tilde{\kappa}_{-N,1}+\tilde{\kappa}_{-N,2} & \tilde{\kappa}_{-N+1,1}+\tilde{\kappa}_{-N+1,2} & \cdots & \tilde{\kappa}_{0,1}+\tilde{\kappa}_{0,2} & \cdots & \tilde{\kappa}_{N-1,1}+\tilde{\kappa}_{N-1,2} & \tilde{\kappa}_{N,1}+\tilde{\kappa}_{N,2} \\ \tilde{\kappa}_{-N,1}-\tilde{\kappa}_{-N,2} & \tilde{\kappa}_{-N+1,1}-\tilde{\kappa}_{-N+1,2} & \cdots & \tilde{\kappa}_{0,1}-\tilde{\kappa}_{0,2} & \cdots & \tilde{\kappa}_{N-1,1}-\tilde{\kappa}_{N-1,2} & \tilde{\kappa}_{N,1}-\tilde{\kappa}_{N,2} \end{bmatrix},$$

such that $a_+ = \sqrt{\frac{1}{2}} \dfrac{(\tilde{\kappa}_{-N,1}+\tilde{\kappa}_{-N,2})s_{-N,+}+\cdots+(\tilde{\kappa}_{0,1}+\tilde{\kappa}_{0,2})s_{0,+}+\cdots+(\tilde{\kappa}_{N,1}+\tilde{\kappa}_{N,2})s_{N,+}}{i(\omega-\tilde{\omega}_+)}$ and

$a_- = \sqrt{\frac{1}{2}} \dfrac{(\tilde{\kappa}_{-N,1}-\tilde{\kappa}_{-N,2})s_{-N,+}+\cdots+(\tilde{\kappa}_{0,1}-\tilde{\kappa}_{0,2})s_{0,+}+\cdots+(\tilde{\kappa}_{N,1}-\tilde{\kappa}_{N,2})s_{N,+}}{i(\omega-\tilde{\omega}_-)}$. The discrete output channels can be formulated by using conservation of energy and time reversal symmetry, and they are expressed as $[s_-] = C[s_+] + K_\Gamma T \begin{bmatrix} a_+ \\ a_- \end{bmatrix}$, where $[s_-] = \begin{bmatrix} s_{-N,-} & s_{-N+1,-} & \cdots & s_{0,-} & \cdots & s_{N-1,-} & s_{N,-} \end{bmatrix}^T$ and C is the nonresonant scattering matrix. For the M = N+1 gap, it is then written explicitly as

$$\begin{bmatrix} s_{-N,-} \\ s_{-N+1,-} \\ \vdots \\ s_{0,-} \\ \vdots \\ s_{N-1,-} \\ s_{N,-} \end{bmatrix} = C[s_+] + \sqrt{\frac{1}{2}} \begin{bmatrix} \tilde{\kappa}_{-N,1} + \tilde{\kappa}_{-N,2} \\ \tilde{\kappa}_{-N+1,1} + \tilde{\kappa}_{-N+1,2} \\ \vdots \\ \tilde{\kappa}_{0,1} + \tilde{\kappa}_{0,2} \\ \vdots \\ \tilde{\kappa}_{N-1,1} + \tilde{\kappa}_{N-1,2} \\ \tilde{\kappa}_{N-1,1} + \tilde{\kappa}_{N-1,2} \end{bmatrix} a_+ + \sqrt{\frac{1}{2}} \begin{bmatrix} \tilde{\kappa}_{-N,1} - \tilde{\kappa}_{-N,2} \\ \tilde{\kappa}_{-N+1,1} - \tilde{\kappa}_{-N+1,2} \\ \vdots \\ \tilde{\kappa}_{0,1} - \tilde{\kappa}_{0,2} \\ \vdots \\ \tilde{\kappa}_{N-1,1} - \tilde{\kappa}_{N-1,2} \\ \tilde{\kappa}_{N-1,1} - \tilde{\kappa}_{N-1,2} \end{bmatrix} a_-, \quad (S3)$$

showing the radiation channels of $a_{+/-}$ are the superpositions of the radiations from $a_{1/2}$. For any given n mirror symmetric pair, as shown in Fig. S1(b), the system symmetry requires $|\tilde{\kappa}_{+n,1}| = |\tilde{\kappa}_{-n,2}|$ and $|\tilde{\kappa}_{+n,2}| = |\tilde{\kappa}_{-n,1}|$ for $a_{1/2}$ propagating in the opposite directions. Since the pair from each mode should have the polarizations pointing either in the same clockwise or counterclockwise direction to ensure continuity, the symmetry further requires two pairs carry opposite polarization directions, or $\tilde{\kappa}_{+n,1} = -\tilde{\kappa}_{-n,2}$ and $\tilde{\kappa}_{+n,2} = -\tilde{\kappa}_{-n,1}$. Eq. (S3) now can be simplified to:

$$\begin{bmatrix} s_{-N,-} \\ s_{-N+1,-} \\ \vdots \\ s_{0,-} \\ \vdots \\ s_{N-1,-} \\ s_{N,-} \end{bmatrix} = C[s_+] + \sqrt{\frac{1}{2}} \begin{bmatrix} \tilde{\kappa}_{-N} - \tilde{\kappa}_N \\ \tilde{\kappa}_{-N+1} - \tilde{\kappa}_{N-1} \\ \vdots \\ 0 \\ \vdots \\ -(\tilde{\kappa}_{-N+1} - \tilde{\kappa}_{N-1}) \\ -(\tilde{\kappa}_{-N} - \tilde{\kappa}_N) \end{bmatrix} a_+ + \sqrt{\frac{1}{2}} \begin{bmatrix} \tilde{\kappa}_{-N} + \tilde{\kappa}_N \\ \tilde{\kappa}_{-N+1} + \tilde{\kappa}_{N-1} \\ \vdots \\ 2\tilde{\kappa}_0 \\ \vdots \\ \tilde{\kappa}_{-N+1} + \tilde{\kappa}_{N-1} \\ \tilde{\kappa}_{-N} + \tilde{\kappa}_N \end{bmatrix} a_-, \quad (S4)$$

where the 1/2 subscripts have been dropped.

We then explicitly formulate the diffraction orders. By considering only one single incidence port q such that $[s_+] = \begin{bmatrix} \cdots & 0 & s_{q,+} & 0 & \cdots \end{bmatrix}^T$, the supermode amplitudes are $a_+ = \sqrt{\frac{1}{2}} \frac{(\tilde{\kappa}_q - \tilde{\kappa}_{-q}) s_{q,+}}{i(\omega - \tilde{\omega}_+)}$ and $a_- = \sqrt{\frac{1}{2}} \frac{(\tilde{\kappa}_q + \tilde{\kappa}_{-q}) s_{q,+}}{i(\omega - \tilde{\omega}_-)}$. The n mirror symmetric diffraction pair thus is:

$$\frac{s_{-n,-}}{s_{q,+}} = \tilde{c}_{-n} - \frac{1}{2}\frac{(\tilde{\kappa}_n - \tilde{\kappa}_{-n})(\tilde{\kappa}_q - \tilde{\kappa}_{-q})}{i(\omega - \tilde{\omega}_+)} + \frac{1}{2}\frac{(\tilde{\kappa}_n + \tilde{\kappa}_{-n})(\tilde{\kappa}_q + \tilde{\kappa}_{-q})}{i(\omega - \tilde{\omega}_-)},$$

$$\frac{s_{+n,-}}{s_{q,+}} = \tilde{c}_n + \frac{1}{2}\frac{(\tilde{\kappa}_n - \tilde{\kappa}_{-n})(\tilde{\kappa}_q - \tilde{\kappa}_{-q})}{i(\omega - \tilde{\omega}_+)} + \frac{1}{2}\frac{(\tilde{\kappa}_n + \tilde{\kappa}_{-n})(\tilde{\kappa}_q + \tilde{\kappa}_{-q})}{i(\omega - \tilde{\omega}_-)},$$

(S5)

where $\tilde{c}_{\pm n}$ are the complex non-resonant scattering coefficients. It is seen from Eq. (S5) that the pairs from $a_+$ and $a_-$ are always $\pi$ out of phase and in phase.

### C. Generalization of the radiation channels of the supermode at the Brillouin zone boundary, X point.

At the X point, Eq. S1 remains the same except $K^T$ now becomes:

$$K_X^T = \begin{bmatrix} \tilde{\kappa}_{-N,1} & \tilde{\kappa}_{-N+1,1} & \cdots & 0 & \cdots & \tilde{\kappa}_{N-1,1} & \tilde{\kappa}_{N,1} \\ \tilde{\kappa}_{-N,2} & \tilde{\kappa}_{-N+1,2} & \cdots & 0 & \cdots & \tilde{\kappa}_{N-1,2} & \tilde{\kappa}_{N,2} \end{bmatrix},$$ where the subscript X stands for the X point

since the normal diffraction is always missing. We diagonalize Eq. S1 in a similar way as

$$\frac{d}{dt}\begin{bmatrix} a_+ \\ a_- \end{bmatrix} = i\begin{bmatrix} \tilde{\omega}_+ & 0 \\ 0 & \tilde{\omega}_- \end{bmatrix}\begin{bmatrix} a_+ \\ a_- \end{bmatrix} + TK_X^T[s_+]$$ with

$$TK_X^T = \sqrt{\frac{1}{2}}\begin{bmatrix} \tilde{\kappa}_{-N,1} + \tilde{\kappa}_{-N,2} & \tilde{\kappa}_{-N+1,1} + \tilde{\kappa}_{-N+1,2} & \cdots & 0 & \cdots & \tilde{\kappa}_{N-1,1} + \tilde{\kappa}_{N-1,2} & \tilde{\kappa}_{N,1} + \tilde{\kappa}_{N,2} \\ \tilde{\kappa}_{-N,1} - \tilde{\kappa}_{-N,2} & \tilde{\kappa}_{-N+1,1} - \tilde{\kappa}_{-N+1,2} & \cdots & 0 & \cdots & \tilde{\kappa}_{N-1,1} - \tilde{\kappa}_{N-1,2} & \tilde{\kappa}_{N,1} - \tilde{\kappa}_{N,2} \end{bmatrix}.$$ After

diagonalization, the supermode amplitudes are

$$a_+ = \sqrt{\frac{1}{2}}\frac{(\tilde{\kappa}_{-N,1} + \tilde{\kappa}_{-N,2})s_{-N,+} + \cdots + 0 + \cdots + (\tilde{\kappa}_{N,1} + \tilde{\kappa}_{N,2})s_{N,+}}{i(\omega - \tilde{\omega}_+)}$$ and

$$a_- = \sqrt{\frac{1}{2}}\frac{(\tilde{\kappa}_{-N,1} - \tilde{\kappa}_{-N,2})s_{-N,+} + \cdots + 0 + \cdots + (\tilde{\kappa}_{N,1} - \tilde{\kappa}_{N,2})s_{N,+}}{i(\omega - \tilde{\omega}_-)}.$$ Eq. (S4) is applied equally well to

the X point except $s_{0,-}$ is always zero. Therefore, for the M = N+1 gap, it is rewritten as:

$$\begin{bmatrix} s_{-N,-} \\ s_{-N+1,-} \\ \vdots \\ s_{0,-} \\ \vdots \\ s_{N-1,-} \\ s_{N,-} \end{bmatrix} = C[s_+] + \sqrt{\frac{1}{2}}\begin{bmatrix} \tilde{\kappa}_{-N} - \tilde{\kappa}_N \\ \tilde{\kappa}_{-N+1} - \tilde{\kappa}_{N-1} \\ \vdots \\ 0 \\ \vdots \\ -(\tilde{\kappa}_{-N+1} - \tilde{\kappa}_{N-1}) \\ -(\tilde{\kappa}_{-N} - \tilde{\kappa}_N) \end{bmatrix}a_+ + \sqrt{\frac{1}{2}}\begin{bmatrix} \tilde{\kappa}_{-N} + \tilde{\kappa}_N \\ \tilde{\kappa}_{-N+1} + \tilde{\kappa}_{N-1} \\ \vdots \\ 0 \\ \vdots \\ \tilde{\kappa}_{-N+1} + \tilde{\kappa}_{N-1} \\ \tilde{\kappa}_{-N} + \tilde{\kappa}_N \end{bmatrix}a_-.$$

(S6)

The n mirror symmetric pair for $a_{+/-}$ still follows Eq. (S5).

### D. FDTD results of 1D SiO$_2$/Au photonic crystals (PhCs)

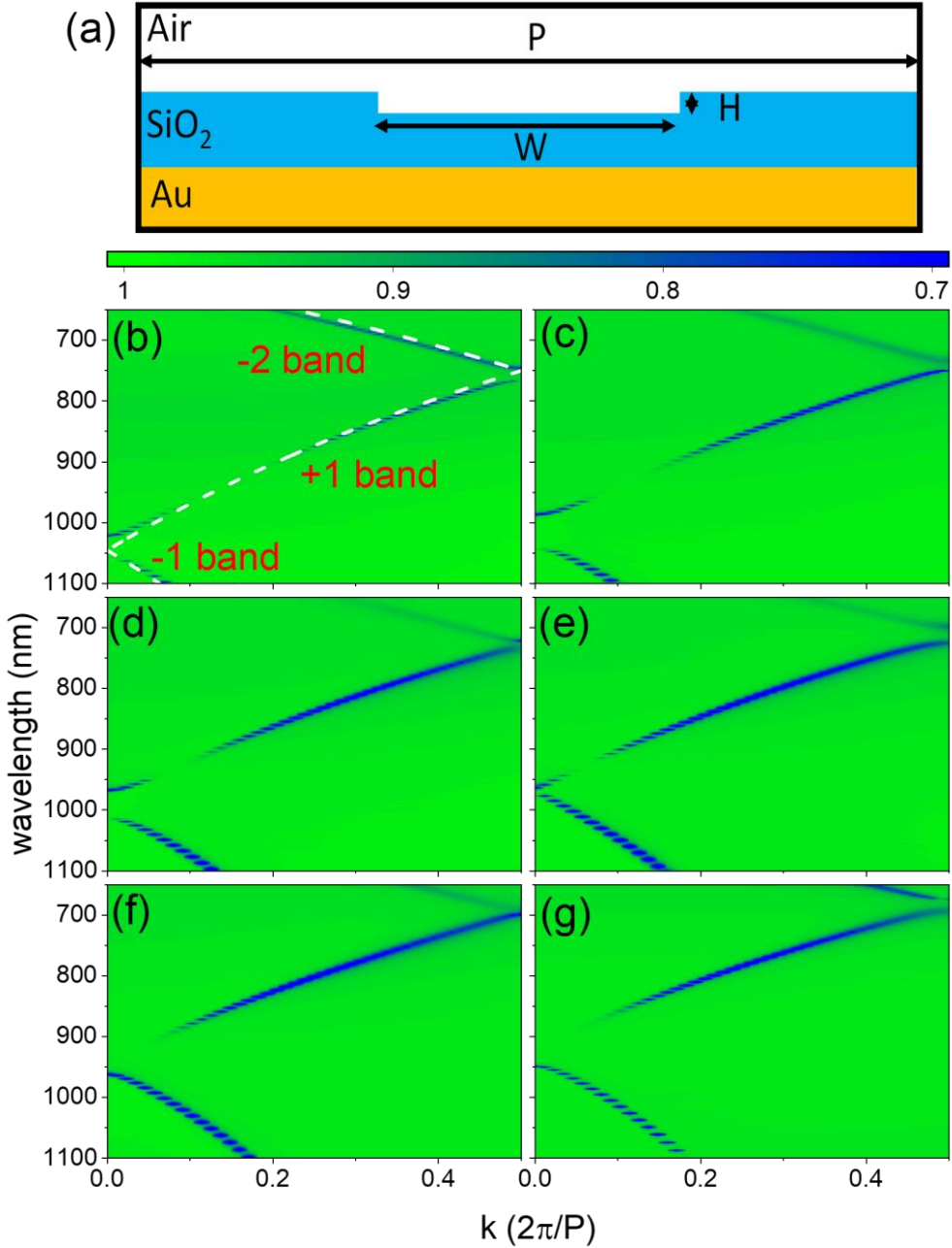

Fig. S2. (a) The FDTD unit cell of the PhC. The simulated TE-polarized k- and λ-resolved total reflectivity mappings of PhCs with W = (b) 100, (c) 225, (d) 350, (e) 475, (f) 600, and (g) 725 nm taken along the Γ-X direction. The white dashed lines are calculated by using the phase-matching equation, indicating the $n_B = \pm 1$ and -2 photonic band are present. At the Γ and X points at k = 0 and 0.5, two energy band gaps are found, featuring two dark and bright modes located above or below the gap. Particularly, at Γ point, a symmetry protected quasi-BIC is observed at either above or below the gap. On the other hand, an accidentally BIC is observed along the +1 band.

Fig. S2(a) shows the unit cell of the PhCs, which has 400 nm thick $SiO_2$ coated on Au surface with the period P and the groove height H being set at 900 nm and 200 nm whereas the groove width W varied from 100 and 725 nm with a step size of 125 nm. The corresponding TE-polarized k-resolved total reflectivity mappings are shown in Fig S2(b) – (f), showing the dispersive ±1 and -2 photonic bands, which follow the phase-matching equation given as $\left(\frac{n_D}{\lambda}\right)^2 = \left(\frac{n_D \sin\theta}{\lambda} + \frac{n_B}{P}\right)^2$, where $n_D$ is the refractive index of $SiO_2$. The calculations are superimposed in Fig 2(b), and we see the $n_B = \pm 1$ photonic bands cross at the Γ point and the $n_B = +1$ and -2 bands cross at the X point, yielding two energy band gaps at λ = 930 – 1030 nm and 700 – 770 nm at the Γ and X points. At the Γ points of all mappings, by visual examination, we find a symmetry protected quasi-BIC is always present, and it is located on the -1 band for W = 100 – 475 nm but flips to the +1 band when W increases further. At the same time, accidental quasi-BICs are also found along the +1 band for all PhCs. We will determine the Zak phases of the +1 photonic band and focus on the supermodes located on the +1 band at the Γ and X points.

At the Γ point, the supermode supports only one m = 0 radiation channel. On the other hand, there are two m = 0 and 1 channels emitted by the supermode at the X point. The reflectivity spectra of the PhCs taken under normal incidence, i.e., at the Γ point, are illustrated in Fig. S3(a), clearly showing only one single reflectivity dip is present as the bright mode, verifying another supermode is quasi-BIC that does not produce any dip. As quasi-BIC arises solely from $a_+$, we deduce the supermode at the +1 band is $a_-$ for W = 100 – 475 nm but becomes $a_+$ for W = 600 and 725 nm. On the other hand, the reflectivity spectra taken at the X point for all PhCs are shown in Fig. S3(b), showing two bright and dark modes are present.

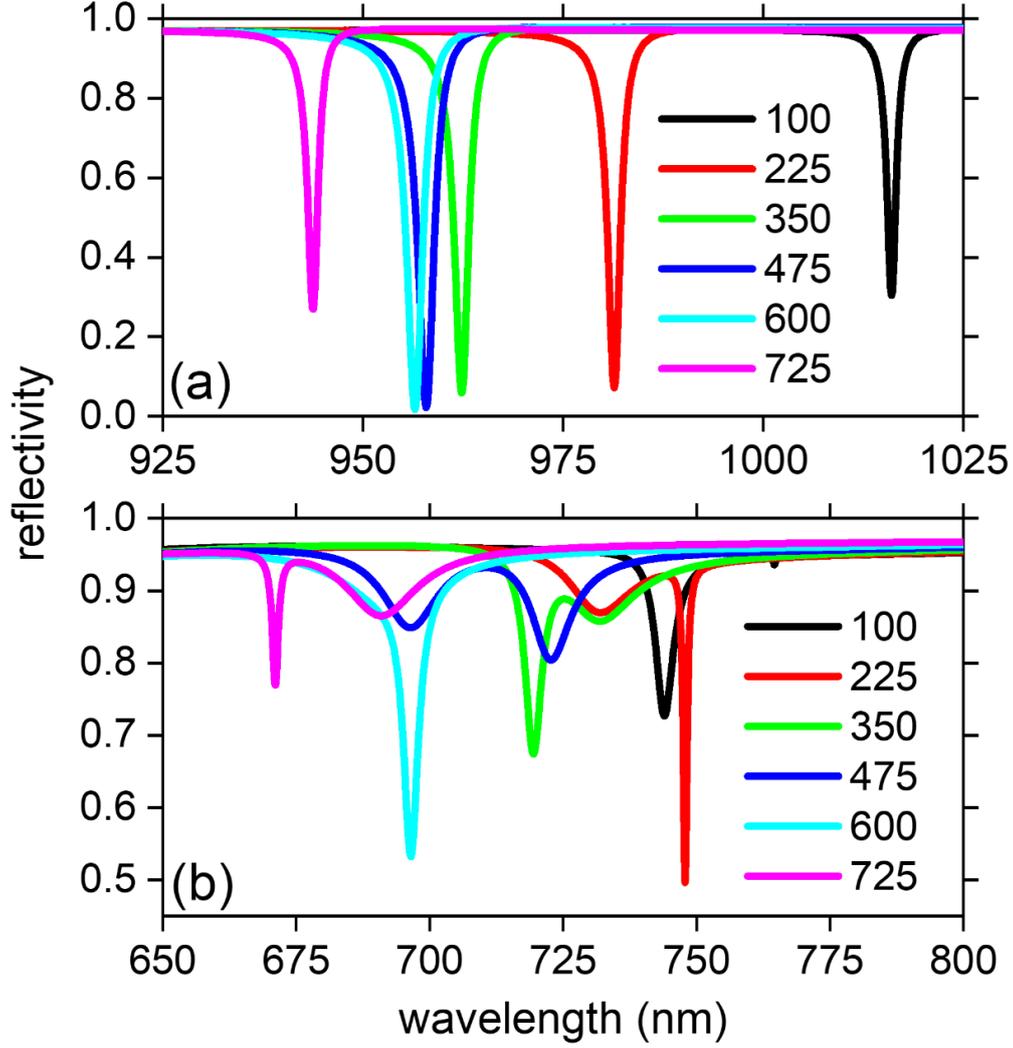

Fig. S3. The TE-polarized total reflectivity spectra of PhCs taken at the (a) Γ and (b) X points for different W. At the Γ point, only one single reflectivity dip is present as the bright mode. On the other hand, at the X point, two bright and dark modes are present.

Two mirror symmetric diffraction and phase spectra are shown in Fig. S4 and they are fitted with $|s_{m=0/1,-}/s_{1,+}|^2$ and $\arg(s_{m=0/1,-}/s_{1,+})$ by using the CMT model. The best fits are displayed as the solid lines and the fitted $\omega_{+/-}$ are tabulated in Table S1, in which the highlights are the supermodes sitting on the +1 photonic band at the zone center and boundary. If the highlights at two regions are either $a_+$ or $a_-$, the Zak phase is 0, but π when they are different. As a result, we conclude the Zak phase of +1 band for W = 100, 225 and 600 nm is π but becomes 0 for W = 350, 475 and 725 nm.

|  |  | 100 nm | 225 nm | 350 nm | 475 nm | 600 nm | 725 nm |
|---|---|---|---|---|---|---|---|
| Γ, zone center | $\omega_+$ (eV) | 1.18 | 1.19 | 1.22 | 1.27 | 1.33 | 1.37 |
|  | $\omega_-$ (eV) | 1.21 | 1.26 | 1.28 | 1.29 | 1.29 | 1.31 |
| X, zone boundary | $\omega_+$ (eV) | 1.62 | 1.65 | 1.72 | 1.78 | 1.79 | 1.79 |
|  | $\omega_-$ (eV) | 1.67 | 1.69 | 1.69 | 1.71 | 1.77 | 1.84 |

Table S1. The FDTD $\omega_{+/-}$ at the Γ and X points for the PhCs with different W. The highlights are the supermodes located at the +1 photonic band. If the highlights at the Γ and X points are both $a_+$ or $a_-$, the Zak phase is 0. If not, the Zak phase is π.

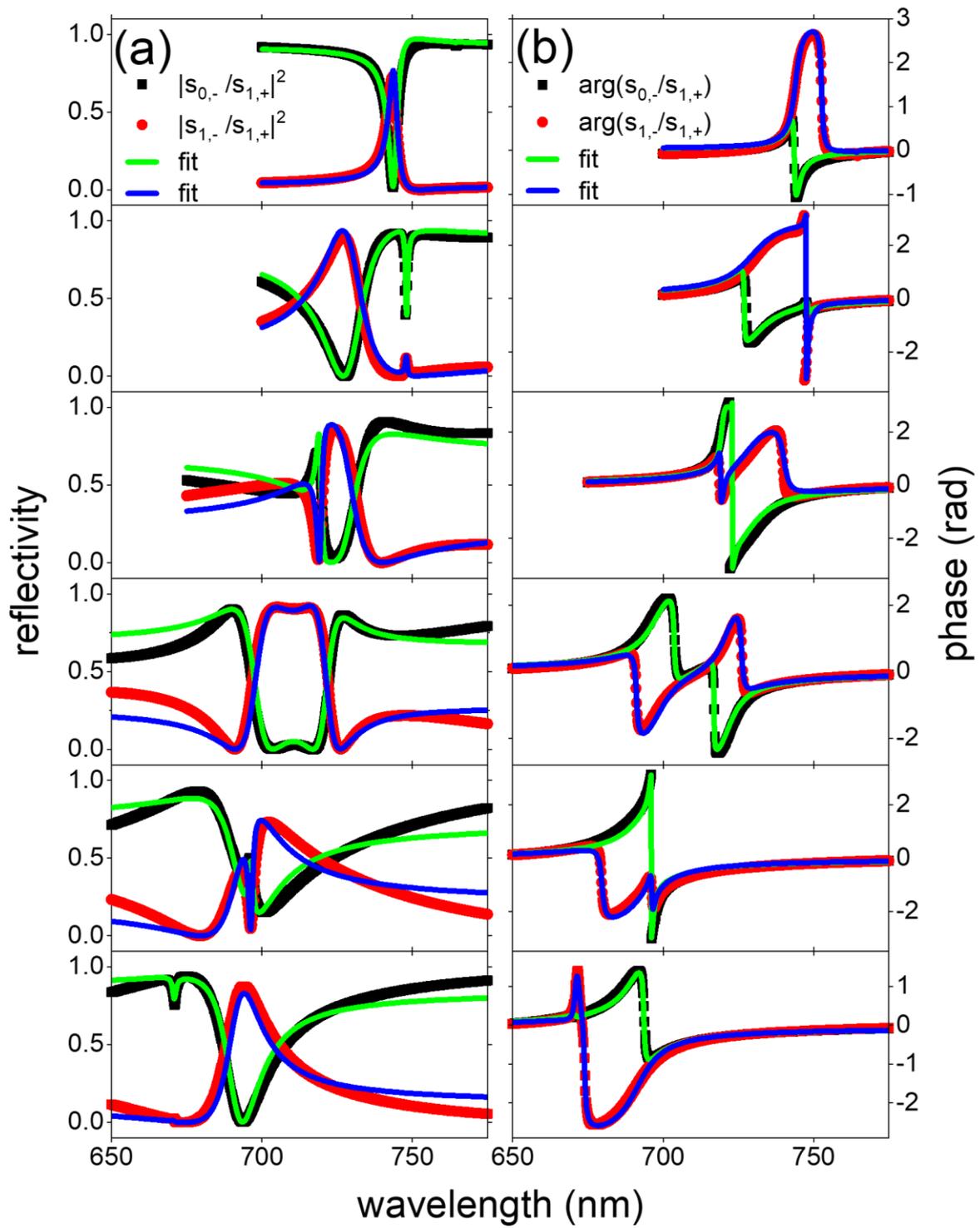

Fig. S4. At the X point, two TE-polarized mirror symmetric m = 0 (black square) and 1 (red circle) (a) reflectivity and (b) phase spectra of the PhCs for W = 100 (top) to 725 (bottom). The green and blue solid lines are the best fits determined by CMT.

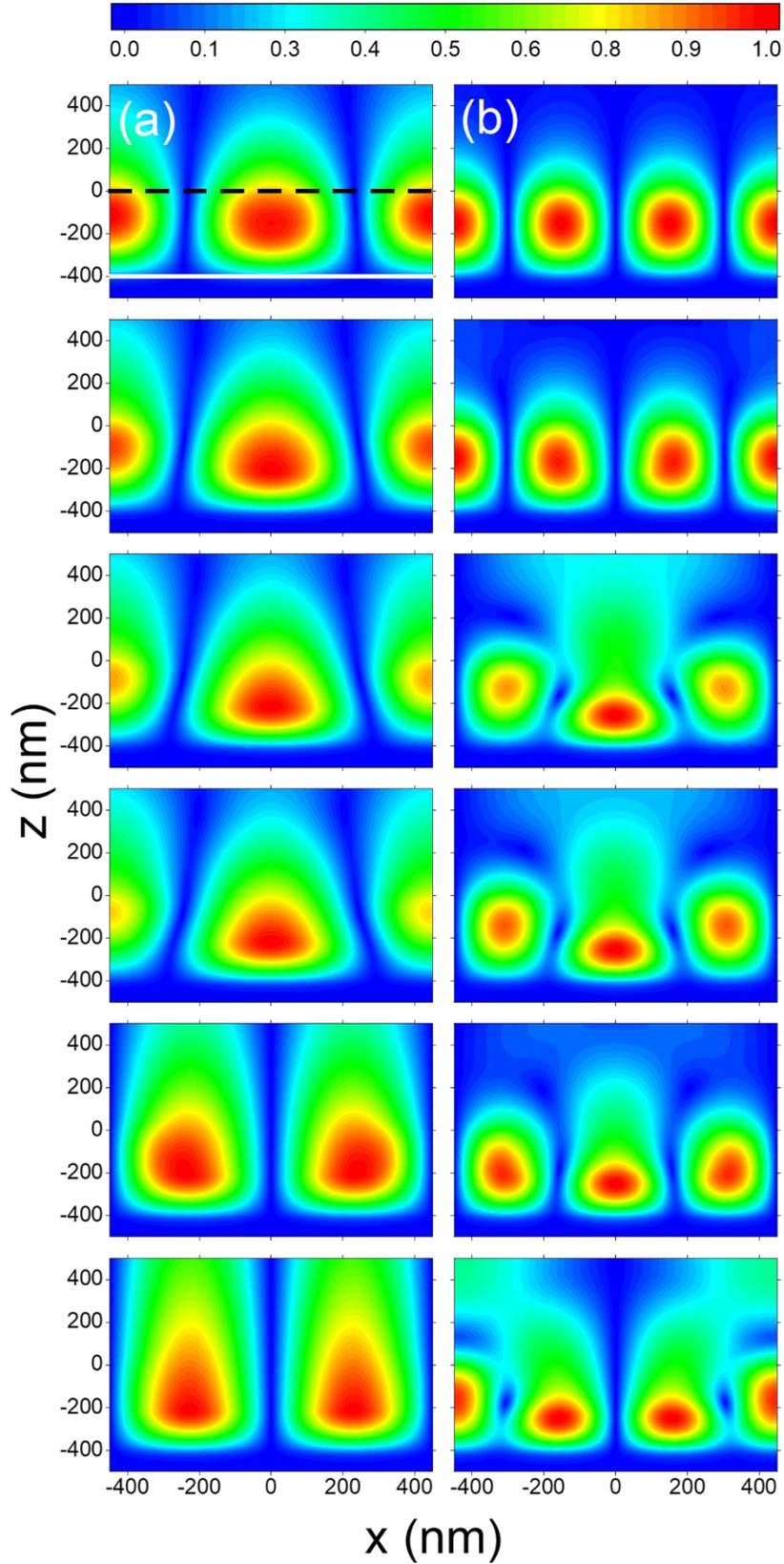

Fig. S5. The FDTD simulated $|E_z|$ near-field patterns of the PhCs for different W taken at the (a) Γ and (b) X points, showing their field symmetries are the same for W = 350, 475 and 725 nm but different for W = 100, 225, and 600 nm. The SiO$_2$ film is shown between z = 0 (black dashed line) and -400 nm (white solid line).

To verify the Zak phases, we have simulated the $|E_z|$ near-field intensity profiles at the Γ and X points of the +1 band in Fig. S5(a) and (b) for different W. At the Γ point, we see the $|E_z|$ profiles are even with respect to the groove center for W = 100 – 475 nm but odd for W = 600 and 725 nm. On the other hand, at the X point, the field symmetries are even for W = 350 – 600 nm but are odd for 100, 225 and 725 nm. Therefore, the comparison between the field symmetries at two points indicates the Zak phases are π for W = 100, 225, and 600 nm and 0 for W = 350, 475 and 725 nm, and they agree very well with earlier CMT results.

### E.  Schematic of the Fourier space optical microscope for angle- and wavelength resolved diffraction mapping and common path interferometry

Fig. S6 shows the schematic of the Fourier space optical microscope. Briefly, a broadband supercontinuum laser from a nonlinear photonic crystal fiber is collimated and then passed through a set of linear polarizers, wave plates, and lenses before being focused onto the back focal plane (BFP) of a 100X objective lens (OB) with numerical aperture = 0.9. The light exiting from the objective lens is then a collimated beam with well-defined linear polarization. In addition, by displacing the focused spot across the BFP of the objective lens using a motorized translation stage, the incident polar angle $\theta_i$ of the collimated beam onto the sample can be varied following $\sin\theta_i = d/f$, where $d$ is the distance between the focused spot and the optical axis of the BFP and $f$ is the focal length of the objective lens. In addition, the azimuth angle $\phi$ can be varied by a motorized rotation sample stage to align the incident plane to the Γ-X direction of the PmC. The diffractions from the PmC are then collected by the same objective lens and are routed through a set of lens system so that the diffraction orders are projected onto the Fourier space. By placing an aperture at the Fourier space to filter out the desired diffraction order, its intensity and phase spectra can be measured by a spectrometer-based CCD detector and a common path interferometer [5].

To perform common path interferometry, the 45° linearly polarized collimated beam with the Jones vector given as $\frac{1}{\sqrt{2}}\begin{bmatrix}1\\1\end{bmatrix}$ is incident on the PmC. The diffraction order from the PmC after the aperture can be formulated as: $J_{PmC} = \begin{bmatrix} |r_{TM}|e^{i\theta_{TM}} & 0 \\ 0 & |r_{TE}|e^{i\theta_{TE}} \end{bmatrix}$, where $r_{TM,TE}$ and $\theta_{TM,TE}$ are the magnitudes and phases for TM- and TE-polarizations. The diffraction passes through a quarter wave plate with the fast axis being placed at 45º with respect to the incident plane and a motorized rotatable analyzer with angle ξ, which are given as

$$J_{analyzer(\xi)} = \begin{bmatrix} \cos^2\xi & \sin\xi\cos\xi \\ \sin\xi\cos\xi & \sin^2\xi \end{bmatrix} \text{ and } J_{QWP(45°)} = \frac{1}{2}\begin{bmatrix} 1-i & 1+i \\ 1+i & 1-i \end{bmatrix}.$$

The output vector is $J_{analyzer(\xi)} J_{QWP(45°)} J_{PmC} \frac{1}{\sqrt{2}}\begin{pmatrix}1\\1\end{pmatrix}$. After some formulations, the intensities for different ξ = 0º, ±45º, and 90º can be written as:

$$R_0(\lambda) = \left| \frac{1}{2}\begin{bmatrix} |r_{TM}|e^{i\varphi} + i|r_{TE}| \\ 0 \end{bmatrix} \right|^2 = \frac{1}{4}\left(|r_{TM}|^2 + |r_{TE}|^2 + 2|r_{TM}||r_{TE}|\sin\varphi\right),$$

$$R_{+45}(\lambda) = \frac{1}{4}\left(|r_{TM}|^2 + |r_{TE}|^2 + 2|r_{TM}||r_{TE}|\cos\varphi\right), R_{-45}(\lambda) = \frac{1}{4}\left(|r_{TM}|^2 + |r_{TE}|^2 - 2|r_{TM}||r_{TE}|\cos\varphi\right),$$ and

$$R_{90} = \frac{1}{4}\left(|r_{TM}|^2 + |r_{TE}|^2 - 2|r_{TM}||r_{TE}|\sin\varphi\right),$$ where $\varphi = \theta_{TM} - \theta_{TE}$. Therefore, the phase difference between TM- and TE- polarized diffractions can be calculated by: $\tan\varphi(\lambda) = \frac{R_0(\lambda) - R_{90}(\lambda)}{R_{+45}(\lambda) - R_{-45}(\lambda)}$.

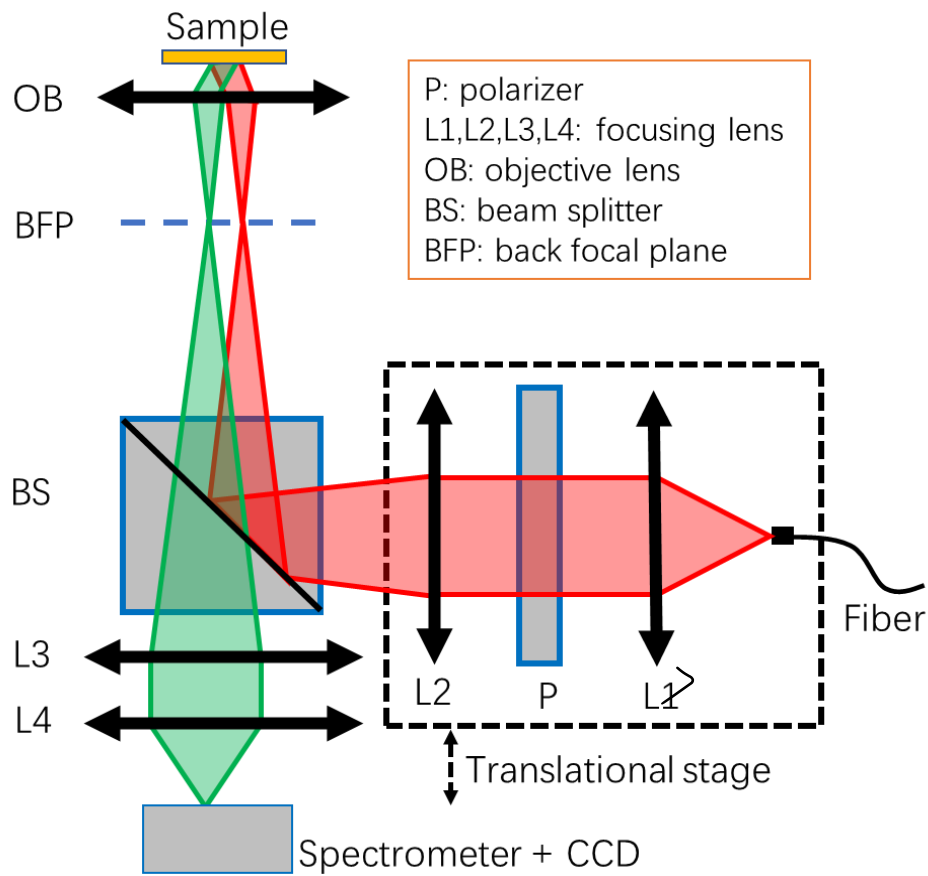

Fig. S6. The schematic of the Fourier optical microscope.